\documentclass[twocolumn]{aastex701}

\usepackage{lipsum}
\usepackage{comment}
\usepackage{amsmath}
\usepackage{graphicx}
\usepackage{orcidlink}
\usepackage[utf8]{inputenc}

\begin{document}

\title{Radio Quasars in the Cosmic Dawn: A Spectroscopic Sample from the DESI Survey} 

\author[0009-0009-8274-441X]{Anniek J. Gloudemans}
\affiliation{NSF NOIRLab, Gemini Observatory, 
670 N A'ohoku Place
Hilo, HI 96720, USA; anniek.gloudemans@noirlab.edu}
\email{anniek.gloudemans@noirlab.edu}  

\author[0000-0001-5287-4242]{Jinyi Yang}
\affiliation{Department of Astronomy, University of Michigan, 1085 S. University Ave., Ann Arbor, MI 48109, USA}\email{} 

\author[0000-0002-5896-6313]{David M. Alexander}
\affiliation{Centre for Extragalactic Astronomy, Department of Physics, Durham University, South Road, Durham, DH1 3LE, UK}\email{} 

\author[0000-0002-4928-4003]{Arjun Dey}
\affiliation{NSF NOIRLab, 950 N. Cherry Avenue, Tucson, AZ 85719, USA}\email{}  

\author[0000-0003-3310-0131]{Xiaohui Fan}
\affiliation{Steward Observatory, University of Arizona, 933 North Cherry Avenue, Tucson, AZ 85721}\email{} 

\author[0000-0003-1251-532X]{Victoria A. Fawcett}
\affiliation{European Southern Observatory (ESO), Karl-Schwarzschild-Straße 2, 85748 Garching bei München, Germany}
\affiliation{School of Mathematics, Statistics and Physics, Newcastle University, NE1 7RU, UK}
\email{}  

\author[0000-0003-4223-1117]{Martin J. Hardcastle}
\affiliation{Centre for Astrophysics Research, Department of Physics, Astronomy and Mathematics, University of Hertfordshire, College Lane, Hatfield AL10 9AB, UK}\email{}  

\author[0000-0002-0000-2394]{Stephanie Juneau}
\affiliation{NSF NOIRLab, 950 N. Cherry Avenue, Tucson, AZ 85719, USA}\email{}  

\author[0000-0003-2884-7214]{George K. Miley}
\affiliation{Leiden Observatory, Leiden University, PO Box 9513, 2300 RA Leiden, The Netherlands}\email{}  

\author[0000-0003-0487-6651]{Leah K. Morabito}
\affiliation{Centre for Extragalactic Astronomy, Department of Physics, Durham University, South Road, Durham DH1 3LE, UK}\affiliation{Institute for Computational Cosmology, Department of Physics, Durham University, South Road, Durham DH1 3LE, UK}\email{}  

\author[]{Adam D. Myers}
\affiliation{Department of Physics and Astronomy, University of Wyoming, Laramie, WY 82071, USA}\email{}  

\author[0000-0002-5854-7426]{Swayamtrupta Panda}
\affiliation{International Gemini Observatory/NSF NOIRLab, Casilla 603, La Serena, Chile}\email{} 

\author[0000-0001-8887-2257]{Huub J. A. R\"{o}ttgering}
\affiliation{Leiden Observatory, Leiden University, PO Box 9513, 2300 RA Leiden, The Netherlands}\email{}  

\author[]{Timothy Shimwell}
\affiliation{Leiden Observatory, Leiden University, PO Box 9513, 2300 RA Leiden, The Netherlands}\affiliation{ASTRON, Netherlands Institute for Radio Astronomy, Oude Hoogeveensedijk 4, 7991 PD, Dwingeloo, The Netherlands}\email{}  

\author[0000-0001-9708-253X]{Daniel J. B. Smith}
\affiliation{Centre for Astrophysics Research, Department of Physics, Astronomy and Mathematics, University of Hertfordshire, College Lane, Hatfield AL10 9AB, UK}\email{}  

\author[0000-0002-7633-431X]{Feige Wang}
\affiliation{Department of Astronomy, University of Michigan, 1085 S. University Ave., Ann Arbor, MI 48109, USA}\email{}  

\author[]{J.~Aguilar}
\affiliation{Lawrence Berkeley National Laboratory, 1 Cyclotron Road, Berkeley, CA 94720, USA}
\email{}

\author[0000-0001-6098-7247]{S.~Ahlen}
\affiliation{Department of Physics, Boston University, 590 Commonwealth Avenue, Boston, MA 02215, USA}
\email{}

\author[0000-0001-9712-0006]{D.~Bianchi}
\affiliation{Dipartimento di Fisica ``Aldo Pontremoli'', Università degli Studi di Milano, Via Celoria 16, I-20133 Milano, Italy}
\affiliation{INAF-Osservatorio Astronomico di Brera, Via Brera 28, 20122 Milano, Italy}
\email{}

\author[]{D.~Brooks}
\affiliation{Department of Physics \& Astronomy, University College London, Gower Street, London, WC1E 6BT, UK}
\email{}

\author[]{T.~Claybaugh}
\affiliation{Lawrence Berkeley National Laboratory, 1 Cyclotron Road, Berkeley, CA 94720, USA}
\email{}

\author[0000-0002-2169-0595]{A.~Cuceu}
\affiliation{Lawrence Berkeley National Laboratory, 1 Cyclotron Road, Berkeley, CA 94720, USA}
\email{}

\author[0000-0002-1769-1640]{A.~de la Macorra}
\affiliation{Instituto de Física, Universidad Nacional Autónoma de México, Circuito de la Investigación Científica, Ciudad Universitaria, Cd. de México C.P. 04510, México}
\email{}

\author[]{P.~Doel}
\affiliation{Department of Physics \& Astronomy, University College London, Gower Street, London, WC1E 6BT, UK}
\email{}

\author[0000-0003-4992-7854]{S.~Ferraro}
\affiliation{Lawrence Berkeley National Laboratory, 1 Cyclotron Road, Berkeley, CA 94720, USA}
\affiliation{University of California, Berkeley, 110 Sproul Hall \#5800, Berkeley, CA 94720, USA}
\email{}

\author[0000-0002-2890-3725]{J.~E.~Forero-Romero}
\affiliation{Departamento de Física, Universidad de los Andes, Cra. 1 No. 18A-10, Edificio Ip, CP 111711, Bogotá, Colombia}
\affiliation{Observatorio Astronómico, Universidad de los Andes, Cra. 1 No. 18A-10, Edificio H, CP 111711 Bogotá, Colombia}
\email{}

\author[0000-0001-9632-0815]{E.~Gaztañaga}
\affiliation{Institut d'Estudis Espacials de Catalunya (IEEC), c/ Esteve Terradas 1, Edifici RDIT, Campus PMT-UPC, 08860 Castelldefels, Spain}
\affiliation{Institute of Space Sciences, ICE-CSIC, Campus UAB, Carrer de Can Magrans s/n, 08913 Bellaterra, Barcelona, Spain}
\email{}

\author[0000-0003-3142-233X]{S.~Gontcho A Gontcho}
\affiliation{University of Virginia, Department of Astronomy, Charlottesville, VA 22904, USA}
\email{}

\author[]{G.~Gutierrez}
\affiliation{Fermi National Accelerator Laboratory, PO Box 500, Batavia, IL 60510, USA}
\email{}

\author[0000-0003-1197-0902]{C.~Hahn}
\affiliation{Institute of Cosmology and Gravitation, University of Portsmouth, Dennis Sciama Building, Portsmouth, PO1 3FX, UK}
\email{}

\author[0000-0002-6024-466X]{M.~Ishak}
\affiliation{Department of Physics, The University of Texas at Dallas, 800 W. Campbell Rd., Richardson, TX 75080, USA}
\email{}

\author[0000-0003-0201-5241]{R.~Joyce}
\affiliation{NSF NOIRLab, 950 N. Cherry Ave., Tucson, AZ 85719, USA}
\email{}

\author[]{R.~Kehoe}
\affiliation{Department of Physics, Southern Methodist University, 3215 Daniel Avenue, Dallas, TX 75275, USA}
\email{}

\author[0000-0002-8828-5463]{D.~Kirkby}
\affiliation{Department of Physics and Astronomy, University of California, Irvine, CA 92697, USA}
\email{}

\author[0000-0003-3510-7134]{T.~Kisner}
\affiliation{Lawrence Berkeley National Laboratory, 1 Cyclotron Road, Berkeley, CA 94720, USA}
\email{}

\author[0000-0001-6356-7424]{A.~Kremin}
\affiliation{Lawrence Berkeley National Laboratory, 1 Cyclotron Road, Berkeley, CA 94720, USA}
\email{}

\author[0000-0003-1838-8528]{M.~Landriau}
\affiliation{Lawrence Berkeley National Laboratory, 1 Cyclotron Road, Berkeley, CA 94720, USA}
\email{}

\author[0000-0001-7178-8868]{L.~Le~Guillou}
\affiliation{Sorbonne Université, CNRS/IN2P3, Laboratoire de Physique Nucléaire et de Hautes Energies (LPNHE), FR-75005 Paris, France}
\email{}

\author[0000-0003-1887-1018]{M.~E.~Levi}
\affiliation{Lawrence Berkeley National Laboratory, 1 Cyclotron Road, Berkeley, CA 94720, USA}
\email{}

\author[0000-0003-4962-8934]{M.~Manera}
\affiliation{Departament de Física, Serra Húnter, Universitat Autònoma de Barcelona, 08193 Bellaterra (Barcelona), Spain}
\affiliation{Institut de Física d’Altes Energies (IFAE), The Barcelona Institute of Science and Technology, Edifici Cn, Campus UAB, 08193 Bellaterra (Barcelona), Spain}
\email{}

\author[0000-0002-1125-7384]{A.~Meisner}
\affiliation{NSF NOIRLab, 950 N. Cherry Ave., Tucson, AZ 85719, USA}
\email{}

\author[]{R.~Miquel}
\affiliation{Institució Catalana de Recerca i Estudis Avançats, Passeig de Lluís Companys, 23, 08010 Barcelona, Spain}
\affiliation{Institut de Física d’Altes Energies (IFAE), The Barcelona Institute of Science and Technology, Edifici Cn, Campus UAB, 08193 Bellaterra (Barcelona), Spain}
\email{}

\author[0000-0002-2733-4559]{J.~Moustakas}
\affiliation{Department of Physics and Astronomy, Siena University, 515 Loudon Road, Loudonville, NY 12211, USA}
\email{}

\author[0000-0001-9070-3102]{S.~Nadathur}
\affiliation{Institute of Cosmology and Gravitation, University of Portsmouth, Dennis Sciama Building, Portsmouth, PO1 3FX, UK}
\email{}

\author[0000-0002-0644-5727]{W.~J.~Percival}
\affiliation{Department of Physics and Astronomy, University of Waterloo, 200 University Ave W, Waterloo, ON N2L 3G1, Canada}
\affiliation{Perimeter Institute for Theoretical Physics, 31 Caroline St. North, Waterloo, ON N2L 2Y5, Canada}
\affiliation{Waterloo Centre for Astrophysics, University of Waterloo, 200 University Ave W, Waterloo, ON N2L 3G1, Canada}
\email{}

\author[0000-0001-7145-8674]{F.~Prada}
\affiliation{Instituto de Astrofísica de Andalucía (CSIC), Glorieta de la Astronomía, s/n, E-18008 Granada, Spain}
\email{}

\author[0000-0001-6979-0125]{I.~P\'erez-R\`afols}
\affiliation{Departament de Física, EEBE, Universitat Politècnica de Catalunya, c/Eduard Maristany 10, 08930 Barcelona, Spain}
\email{}

\author[]{A.~Robertson}
\affiliation{NSF NOIRLab, 950 N. Cherry Ave., Tucson, AZ 85719, USA}
\email{}

\author[]{G.~Rossi}
\affiliation{Department of Physics and Astronomy, Sejong University, 209 Neungdong-ro, Gwangjin-gu, Seoul 05006, Republic of Korea}
\email{}

\author[0000-0002-9646-8198]{E.~Sanchez}
\affiliation{CIEMAT, Avenida Complutense 40, E-28040 Madrid, Spain}
\email{}

\author[0000-0002-3569-7421]{E.~F.~Schlafly}
\affiliation{Space Telescope Science Institute, 3700 San Martin Drive, Baltimore, MD 21218, USA}
\email{}

\author[]{D.~Schlegel}
\affiliation{Lawrence Berkeley National Laboratory, 1 Cyclotron Road, Berkeley, CA 94720, USA}
\email{}

\author[0000-0002-3461-0320]{J.~Silber}
\affiliation{Lawrence Berkeley National Laboratory, 1 Cyclotron Road, Berkeley, CA 94720, USA}
\email{}

\author[]{D.~Sprayberry}
\affiliation{NSF NOIRLab, 950 N. Cherry Ave., Tucson, AZ 85719, USA}
\email{}

\author[0000-0003-1704-0781]{G.~Tarl\'e}
\affiliation{University of Michigan, 500 S. State Street, Ann Arbor, MI 48109, USA}
\email{}

\author[]{B.~A.~Weaver}
\affiliation{NSF NOIRLab, 950 N. Cherry Ave., Tucson, AZ 85719, USA}
\email{}

\author[0000-0001-5381-4372]{R.~Zhou}
\affiliation{Lawrence Berkeley National Laboratory, 1 Cyclotron Road, Berkeley, CA 94720, USA}
\email{}

\author[0000-0002-6684-3997]{H.~Zou}
\affiliation{National Astronomical Observatories, Chinese Academy of Sciences, A20 Datun Road, Chaoyang District, Beijing 100101, P.~R.~China}
\email{}


\begin{abstract}

We present the largest statistical spectroscopic sample of radio quasars at $4.7\leq z \leq6.7$ to date. These have been selected from a dedicated Dark Energy Survey Instrument (DESI) target program and provide a robust sample to study the influence of radio jets on early quasar evolution. Using low-frequency radio data from the LOFAR Two Metre Sky Survey Data Release 3 (LoTSS-DR3) at 144 MHz, we identify 83 high-$z$ radio quasars, 26 of which are classified as radio-loud (5-15\% of 646 confirmed high-$z$ quasars, defined as $R = F_{5\text{GHz}}/F_{4400\text{\AA}} >10$). With radio luminosities of $\sim10^{25-28}$ W Hz$^{-1}$, the radio emission in all 83 quasars is expected to be jet-dominated rather than powered by star formation. Analysis of their rest-frame ultra-violet (UV) spectra reveals broadly similar composite spectral properties and \ion{C}{4} velocity shifts compared to radio non-detected quasars. The radio-detected and radio-loud quasars exhibit slightly smaller Ly$\alpha$ and \ion{C}{4} equivalent widths on average, however, we find no significant evidence ($\sim1.3\sigma$) for a higher fraction of weak-line quasars (WLQs; $\sim10\pm4$\% versus $\sim4\pm1$\% for radio non-detected quasars). We further find that the broad absorption line (BAL) quasars are predominantly radio-quiet, and no correlation is observed between absorption index and radio-loudness, suggesting that radio emission is not produced by the BAL outflows themselves and are physically distinct phenomena. Our results demonstrate that powerful radio jets were already well-established by $z\sim5$.

\end{abstract}


\section{Introduction} 
\label{sec:intro}

Being the most luminous non-transient sources in our Universe, quasars offer a unique possibility to study supermassive black hole (SMBH) activity and growth across cosmic time. Quasars have now been discovered up to redshift $z\approx7.8$ (e.g., \citealt{Banados2018Natur.553..473B, Yang2020ApJ...897L..14Y, wang2021ApJ...907L...1W, Yang2026A&A...711A.104Y}), with hundreds of quasar discoveries at $z\sim5-6$ thanks to all-sky imaging survey such as PanSTARRS (PS1; \citealt{Chambers2016arXiv161205560C}). It has been established that roughly $\sim$10\% of the optically selected quasar population emits strong radio emission; these have historically been classified as radio-loud\footnote{Quasars are generally classified as radio-loud when $R = F_{5\text{GHz}}/F_{4400\text{\AA}} >10$ in rest-frame \citep{Kellermann1989AJ.....98.1195K}.}. While the physical origin of radio-loud versus radio-quiet quasars has been heavily debated over the past few decades (e.g., \citealt{Kellermann1989AJ.....98.1195K, Ivezic2002AJ....124.2364I, Cirasuolo2003MNRAS.346..447C, Balokovic2012ApJ...759...30B}), recent large sky radio surveys have demonstrated that radio-loud and radio-quiet quasars likely have the same physical origin with both the active galactic nuclei (AGN) and star formation (SF) contributing to their radio emission (e.g., \citealt{Gurkan2019A&A...622A..11G, Macfarlane2021MNRAS.506.5888M, Yue2024MNRAS.529.3939Y}). Radio-loud quasars simply represent the tail-end of the radio luminosity distribution with their radio emission mostly generated by powerful radio jets. These jets regulate BH and galaxy growth via feedback and influence their larger scale environment. Their energetic outflows can both suppress and trigger star formation in their galaxies by heating
the circumgalactic gas and compressing dense clouds (e.g., \citealt{Silk1998A&A...331L...1S, Bower2006MNRAS.370..645B}). Furthermore, radio-loud quasars are frequently found in overdense regions (e.g., \citealt{Wylezalek2013ApJ...769...79W, Overzier2016A&ARv..24...14O}), making them excellent signposts for identifying protoclusters and investigating the early formation of large-scale structure in the Universe (e.g., \citealt{Harikane2019_SR8}).  

The advent of multi-fiber spectrographs such as the Sloan Digital Sky Survey (SDSS; \citealt{York2000AJ....120.1579Y}) and Dark Energy Spectroscopic Instrument (DESI; \citealt{DESI2022AJ....164..207D}) has enabled efficient spectroscopic follow-up of large samples of quasar candidates. These and other surveys have revealed a diverse population of quasars exhibiting a wide range of unusual spectral properties, including broad absorption line quasars (BAL QSOs; e.g., \citealt{Weymann1991ApJ...373...23W, Trump2006ApJS..165....1T}), which show broad blueshifted absorption troughs in the rest-frame ultra-violet (UV) spectra. The BAL phenomenon is quite common in quasars, with reported fractions of $\sim10-20$\% in optically selected samples (e.g., \citealt{Weymann1991ApJ...373...23W, Hewett2003AJ....125.1784H, Trump2006ApJS..165....1T, Gibson2009ApJ...692..758G, Yang2021ApJ...923..262Y}) and fractions as high as $\sim20-40$\% in infrared and radio selected samples (e.g., \citealt{Becker2000ApJ...538...72B, Dai2008ApJ...672..108D, Knigge2008MNRAS.386.1426K, Ganguly2008ApJ...672..102G, Maddox2008MNRAS.386.1605M, Urrutia2009ApJ...698.1095U}). These BAL features are thought to be caused by outflowing gas and quasar winds; however, their unification with non-BAL quasars is still debated. One explanation is that this absorption due to outflowing gas occurs in all quasars, but is only detected for quasars with a specific orientation when the outflowing material intersects our line of sight. Alternatively, BAL quasars could represent a short stage in early quasar evolution with spherically symmetric outflows. Although both orientation and evolution could play a role, some studies have found that the BAL phenomenon is more likely the result of orientation (e.g., \citealt{Elvis2000ApJ...545...63E, Ghosh2007ApJ...661L.139G, Naddaf2023A&A...675A..43N}), while other work supports the idea of an early evolutionary stage (e.g., \citealt{Hazard1984ApJ...282...33H, Becker2000ApJ...538...72B, Bischetti2023ApJ...952...44B}). Radio observations can support the investigation of this question, since their radio spectra and resolved radio jets can give an indication of orientation (e.g., \citealt{Barthel1989ApJ...336..606B, Becker2000ApJ...538...72B, Morabito2017MNRAS.469.1883M}). However, the majority of BAL quasars are radio-quiet (e.g., \citealt{Stocke1992ApJ...396..487S, DiPompeo2011ApJ...743...71D}) and only a few radio-loud BAL quasars (e.g., \citealt{Brotherton1998ApJ...505L...7B, Becker2000ApJ...538...72B}) are suitable probes for orientation via radio morphology. Recent work suggests that their radio emission and absorption mechanisms might be linked to the same underlying process, such as AGN winds, while they are are spatially distinct phenomena (e.g., \citealt{Morabito2019A&A...622A..15M, Petley2024MNRAS.529.1995P}). 

Other intriguing finds in spectroscopic samples are weak-line quasars (WLQs; e.g., \citealt{Diamond-Stanic2009ApJ...699..782D}), which are characterized by their low equivalent width or completely missing broad high-ionization emission lines. Different physical mechanisms have been suggested to explain these quasar spectra, including an undeveloped broad line region (BLR) from young quasars (e.g., \citealt{Hryniewicz2010MNRAS.404.2028H, Plotkin2015ApJ...805..123P}), high Eddington accretion rate causing a soft ionizing continuum (e.g., \citealt{Leighly2007ApJS..173....1L, Laor2011MNRAS.417..681L, Meusinger2014A&A...568A.114M}), and artificial continuum boosting due to relativistic jets (e.g., \citealt{Meusinger2014A&A...568A.114M}). However, previous studies by \cite{Shemmer2009ApJ...696..580S} and \cite{Plotkin2010ApJ...721..562P} concluded that the radio, optical variability, polarization, and X-ray properties of the majority of WLQs are inconsistent with a relativistically beamed jet (BL Lac-like) interpretation.
The radio-loudness of WLQs found in low-$z$ studies is still debated with works finding no statistical difference between the radio-loudness of WLQs and normal quasars (see \citealt{Diamond-Stanic2009ApJ...699..782D, Shemmer2009ApJ...696..580S}) and others finding a high percentage of radio-loud quasars amongst the WLQ population (26\% by \citealt{Meusinger2012A&A...541A..77M}, and 22-37\% by \citealt{Meusinger2014A&A...568A.114M}). 
Furthermore, there is evidence for an increase in the fraction of WLQs at high-$z$, going from 1-6\% at $3<z<5$ \citep{Diamond-Stanic2009ApJ...699..782D} to 14\% at $z>5.6$ \citep{Banados2016ApJS..227...11B}, and some works suggest that this fraction is even higher for radio-loud quasars at high redshift up to 20-40\% \citep{Banados2014AJ....148...14B, Gloudemans2022A&A...668A..27G}. However, these high-$z$ studies are based on small sample sizes and biased by selection effects, and require further investigation.  

Quasars typically exhibit comparable rest-frame UV properties, even though their radio characteristics vary considerably. Their radio emission likely traces longer timescales than the rest-frame UV emission from the broad-line region (especially for extended jets), which can quickly change in response to the accretion state (e.g., \citealt{Ulrich1997ARA&A..35..445U, VandenBerk2004ApJ...601..692V}). However, despite potentially probing different timescales, recent statistical population studies at low redshift using SDSS suggest that the radio-loud quasars display a redder continuum and enhanced [O\textsc{ii}] emission \citep{Arnaudova2024MNRAS.528.4547A}, and preferentially exist at low \ion{C}{4} blueshifts compared to radio-undetected quasars (e.g., \citealt{Richards2011AJ....141..167R, Kratzer2015AJ....149...61K, Rankine2021MNRAS.502.4154R, Jackson2026MNRAS.546ag065J}). Furthermore, \cite{Yue2025MNRAS.537..858Y} found that quasars with the top 20\% most massive SMBHs are 2-3$\times$ more likely to host strong radio jets than similar lower mass quasars. These results imply that radio-loudness is at least partly an intrinsic property of the host system rather than simply being a phase that all quasars go through with equal probability. One of the key questions that remains is why some quasars generate powerful radio jets while others do not. Specifically at high redshift, there seems to be a lack of large extended ($>100$ kpc) radio jets (see \citealt{DeBreuck1999A&A...352L..51D, Miley2008A&ARv..15...67M, Momjian2018ApJ...861...86M}), while abundant at low-$z$, with the most extended radio jet at $z\geq4$ measured to be $66$ kpc in projected length \citep{Gloudemans2025ApJ...980L...8G}. Although selection and observational effects, such as the increase of the cosmic microwave background (CMB) energy density at high-$z$, likely play an important role, it remains unclear when the first radio jets form, how they evolve, and how these influence early galaxy evolution. 

Despite recent efforts to increase the number of radio galaxies and quasars known at high-$z$ (see e.g., \citealt{Banados2015ApJ...804..118B, Saxena2019MNRAS.489.5053S, Gloudemans2022A&A...668A..27G, Ighina2025A&A...698A.158I}), the sample size is modest with only $\gtrsim20$ radio-loud quasars known at $z>5$. This sample is highly non-uniform, with quasars selected using different techniques and spectra from different telescopes and instruments. The newly discovered quasars at $4.8< z < 6.8$ identified by a DESI secondary target program (see \citealt{Yang2023ApJS..269...27Y} and Yang et al. in prep.), together with deep low-frequency radio maps from the LOFAR Two Metre Sky Survey (\citealt{Shimwell2022A&A...659A...1S, Shimwell2026arXiv260215949S}), allow us to build a statistically meaningful and uniform spectroscopic sample of high-$z$ radio quasars. This work aims to investigate the connection between this radio emission and spectral features such as emission line strengths, broad absorption lines, and \ion{C}{4} blueshift (indicative of outflows) to improve our understanding of the influence of radio jets on early quasar evolution. 

\begin{figure*}[ht]
    \centering
    \includegraphics[width=0.80\linewidth]{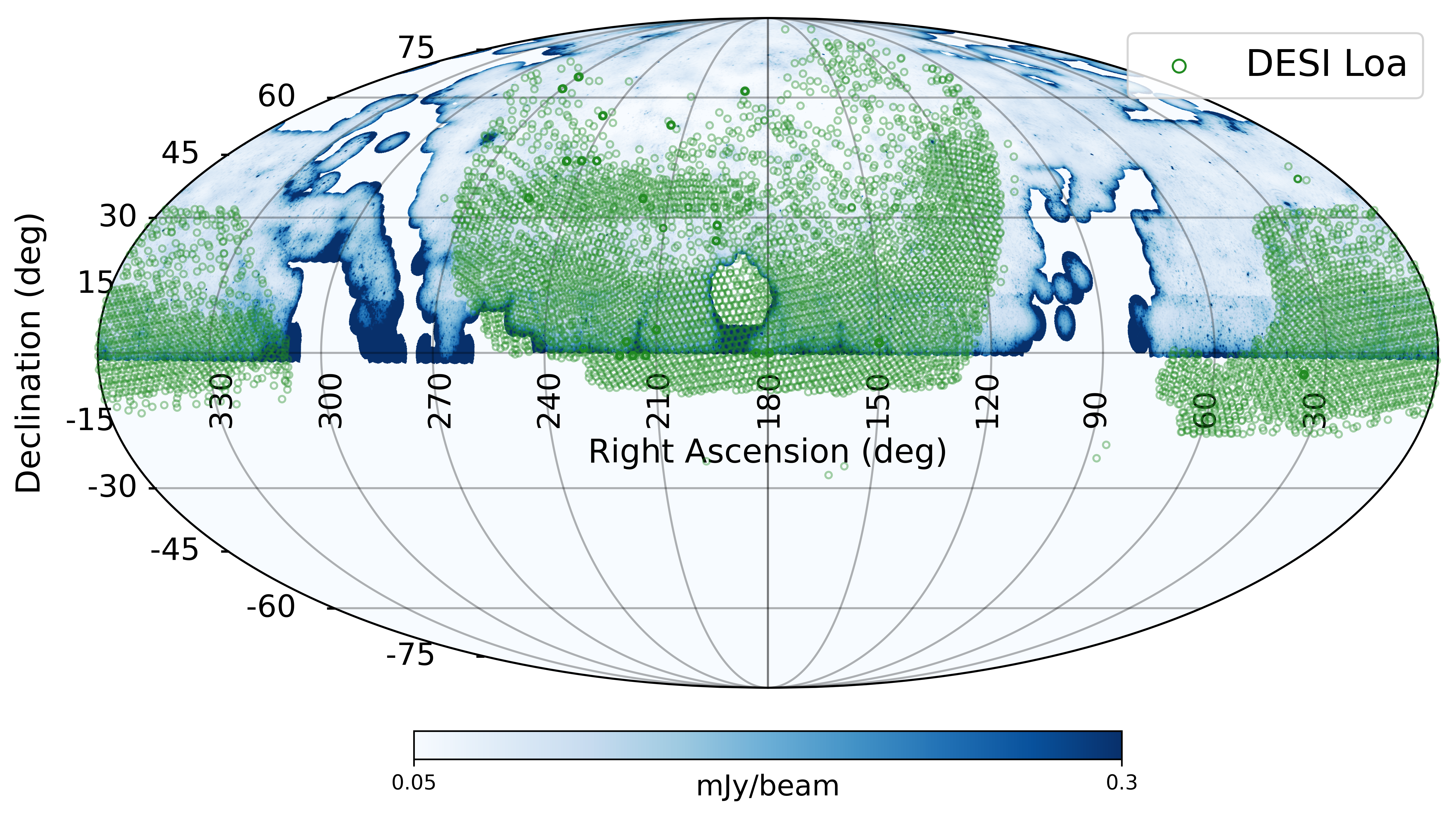}
    \caption{
    The LoTSS-DR3 sky coverage (light blue) and overlap with the DESI Loa tiles (green circles). The LoTSS-DR3 footprint covers 88\% of the Northern sky. The LoTSS RMS noise level (in mJy beam$^{-1}$) is indicated with the colorbar and increases towards the equator, thereby decreasing our sensitivity to radio quasars. Out of the 646 high-$z$ DESI quasars studied in this work, 510 are covered by LoTSS with a median RMS noise level of 85 $\mu$Jy beam$^{-1}$.} 
    \label{fig:lotss_rms}
\end{figure*} 

This paper is structured as follows: in Sect.~\ref{sec:desi_quasar_sample} the DESI $z\gtrsim5$ quasar sample is presented with their radio detections and rest-frame UV properties. In Sect.~\ref{sec:spec_analysis}, the spectral analysis procedure and selection of BAL quasars is explained. The main results comparing the spectral properties of the different samples of high-$z$ (radio) quasars are presented in Sect.~\ref{sec:results}. Finally, the implications of this work are discussed in Sect.~\ref{sec:discussion}, and the results are summarized in Sect.~\ref{sec:results}. Throughout this paper, we use the AB magnitude system and assume a $\Lambda$-CDM cosmology with H$_{0}$= 70 km s$^{-1}$ Mpc$^{-1}$, $\Omega_{M}$ = 0.3, and $\Omega_{\Lambda}$ = 0.7. When performing statistical tests, our threshold for statistical significance is set at a $p$-value of $<0.05$.

\section{DESI $z\gtrsim5$ Quasar Sample}
\label{sec:desi_quasar_sample}

The Dark Energy Spectroscopic Instrument (DESI) is a robotic, fiber-fed, highly multiplexed spectroscopic surveyor operating on the Mayall 4-meter telescope at Kitt Peak National Observatory \citep{DESI2022AJ....164..207D}. It can simultaneously obtain spectra for nearly 5,000 objects over a field of roughly 3 deg$^2$ \citep{DESI2016arXiv161100036D, DESI_Instrument_Design_2016, DESI_Corrector_2024, DESI_Fiber_System_2024}. DESI is currently carrying out an eight-year survey \citep{Schlafly2023AJ....166..259S} covering approximately 17,000 deg$^2$ of the sky.
The First Data Release (DR1, \citealt{DESIcollaboration2025arXiv250314745D}), including spectra for $\sim18$ million targets, is now public. The resulting new constraints on dark energy have been published in \cite{DESI_DR1_cosmology_2024} and \cite{DESI_DR2_cosmology_2025}. The completed survey is expected to spectroscopically confirm 63 million galaxies and quasars \citep{Guy2023AJ....165..144G}. 

The quasar sample investigated in this work originates from a DESI secondary target program (see Sect. 3.2 in \citealt{Myers2023AJ....165...50M}) designed to identify quasars at $4.8< z < 6.8$ \citep{Yang2023ApJS..269...27Y}. These targets have been selected with a color selection technique using primarily photometry from the Legacy Surveys DR9 \citep{dey2019AJ....157..168D}, Pan-STARRS1 survey (PS1; \citealt{Chambers2016arXiv161205560C}), Wide-field Infrared Survey Explorer (WISE; \citealt{wright2010AJ....140.1868W}), and various public J-band surveys, including the UKIRT Hemisphere Survey \citep{Dye2018MNRAS.473.5113D}, UKIRT InfraRed Deep Sky Surveys–Large Area Survey \citep{Lawrence2007}, and VISTA Hemisphere Survey \citep{McMahon2013Msngr.154...35M}. The first year of observations led to the discovery of 400 new quasars at $4.7\lesssim z \lesssim 6.6$, almost doubling the number of known quasars at these redshifts (see \citealt{Yang2023ApJS..269...27Y} for details). Recently, the survey identified an additional 247 quasars at $z\gtrsim4.7$, creating a statistical sample of 647 high-$z$ quasars. This sample is preliminary, as the full sample of the $z\gtrsim5$ quasar survey program will be published in J. Yang et al. (2026 in prep.) when the program completes. We refer to this paper for details on the sample selection and complete analysis of the quasar properties. This paper is based on the spectra from the internal DESI ``Loa'' data release (see \citealt{Guy2023AJ....165..144G} for general information on the pipeline), which includes targets observed by April 2024. To provide a clean sample, the quasar redshifts have been determined using visual inspection. We removed one of the quasars included by \cite{Yang2023ApJS..269...27Y} in this work, named DESI J144444.70+643508.4, since this spectrum was taken as part of the Early Data Release under the Survey Validation \citep{DESI2024AJ....167...62D} and was not included in the Loa data release, making a final dataset of 646 high-$z$ quasar spectra.

\subsection{Radio detections from LoTSS-DR3}
\label{subsec:lotss_radio_detections}

To investigate the radio properties of this high-$z$ quasar sample, we used the LOFAR Two Metre Sky Survey (LoTSS; \citealt{Shimwell2019A&A...622A...1S, Shimwell2022A&A...659A...1S, Shimwell2026arXiv260215949S}), which aims at observing the entire Northern Hemisphere at a radio frequency of 144 MHz and resolution of 6-9\arcsec\ with unprecedented sensitivity of $< 100\, \mu$Jy beam$^{-1}$ (see Fig.~\ref{fig:lotss_rms}). The latest data release DR3 \citep{Shimwell2026arXiv260215949S} covers 79\% (510/646) of our quasars. As demonstrated in Fig.~\ref{fig:lotss_rms}, due to the Northern latitude of LOFAR, its sensitivity decreases toward the equator. The RMS sensitivity typically varies between 70-150 $\mu$Jy beam$^{-1}$ with a median of 85 $\mu$Jy beam$^{-1}$. For LOFAR-detected quasars at $z\sim5$, the observed 144 MHz emission corresponds to a rest-frame frequency of $\sim900$ MHz. Consequently, the radio emission probes different physical components than in comparable low-$z$ samples, with relatively greater sensitivity to compact core/jet emission and reduced sensitivity to extended steep-spectrum lobes. The mix of radio quasar types in the high-redshift sample may therefore differ from that in low-$z$ radio-selected populations.

By performing a nearest-neighbor cross-match between the quasar catalog and LoTSS-DR3 catalog (v1.0) within 1.5\arcsec, we found that 53 out of 510 quasars are detected at 144 MHz with high significance ($>5\sigma$). This cross-matching radius is based on the work of \cite{Macfarlane2021MNRAS.506.5888M} using the LoTSS catalog, which demonstrated that the contamination increases steeply above a radius of 1.5\arcsec. To verify this is the case for our sample as well, we searched for larger-scale radio emission within 10\arcsec, which yielded 21 additional potential radio sources that have not been identified by either the 1.5\arcsec\, cross-matching or our method described below. However, visual inspection revealed that this radio emission is most likely associated with foreground galaxies in the vicinity, and therefore these radio detections are not included in our analysis.  

To be able to also detect faint ($3-5\sigma$ significance) radio emission where the LoTSS catalog is incomplete, we additionally performed two-dimensional Gaussian fitting at the optical location of each quasar\footnote{Performing forced photometry as an alternative results in similar source detections, with only 1 potentially faint radio source missed by the Gaussian fitting routine.}\textsuperscript{,}\footnote{Images have been downloaded from the internal LOFAR archive on 13 Feb 2026.}. First, we created cutouts from the LoTSS-DR3 mosaic of $60\arcsec\times60\arcsec$ around each quasar location. Second, assuming the radio sources are unresolved, we fitted a two-dimensional Gaussian fixed to the beam size and position angle at each source location. The location of the Gaussian peak was allowed to vary by 3 pixels (4.5\arcsec) in each direction. To be detected, the source had to satisfy two criteria, i) its signal-to-noise (S/N) ratio is higher than $3$, and ii) the radio source is within 2\arcsec\, of the optical quasar. We allowed the offset to be slightly larger compared to sources in the LoTSS-DR3 catalog, because the positional uncertainty increases with decreasing S/N. The rms noise in each image was determined by taking the standard deviation using the median absolute deviation of the entire $60\arcsec\times60\arcsec$ cutout image. The peak flux density of each source was determined by taking the maximum flux density value within $2\times$ the beam size of the optical quasar location. Through this process, we identified 85 radio-detected quasars, including 52 quasars that were already detected in the LoTSS-DR3 source catalog. Only 1 quasar from the LoTSS-DR3 catalog (ILTJ012634.54+281218.2 at $z=4.74$) was not detected by our fitting routine due to its low S/N of 3.0. Finally, we visually inspected all candidates and removed 3 low S/N and extended radio detections that did not seem to be associated with the quasar. Our final radio-detected quasar sample includes 83 radio-detected quasars (53 from the catalog and 30 additional sources from our fitting procedure). Their optical cutouts and radio contours are shown in Fig.~\ref{fig:detected_radio_sources}. To account for any systematic flux uncertainties, we added a 10\% flux density error in quadrature to all measured radio flux densities, which is approximately the flux density scale accuracy of the LoTSS survey (see \citealt{Shimwell2022A&A...659A...1S}). We note that we could still be missing very faint radio sources ($<0.3$ mJy beam$^{-1}$), especially when their radio emission is extended. 

\subsection{Other radio surveys}
\label{subsec:other_radio_surveys}

To ensure that no bright radio quasars are missed in our selection, we also investigated the two main other Northern large sky surveys: Faint Images of the Radio Sky at Twenty-cm survey at 1.4 GHz (FIRST; \citealt{Becker1994ASPC...61..165B}) and the Very Large Array Sky Survey at 2-4 GHz (VLASS; \citealt{Lacy2020PASP..132c5001L}). From the full high-$z$ quasar sample, 10 are detected within 1.5\arcsec\, in the FIRST catalog. Only 1 of these (DESI J115838.96-052226.34 at $z=4.93$) is not detected in LoTSS, because it is outside of the LoTSS footprint. We perform the same Gaussian fitting technique as described in Sect.~\ref{subsec:lotss_radio_detections} on the FIRST images, which yields 6 extra source detections, with 5 of them missed in LoTSS. However, all 5 radio detections are low signal-to-noise (S/N$\sim3-4$) and have total flux densities of $0.4-0.5$ mJy.
For the VLASS survey, we obtained quicklook continuum images from the most recent data release, ranging from VLASS 1.1 to 4.1, for each quasar. Again, we performed the Gaussian fitting, which resulted in 32 $>3\sigma$ detections, 14 of which are not detected in LoTSS (with 2 not in the LoTSS region). Therefore, there are 12 potential extra radio quasars in our sample. These are again faint, with a total flux density of $0.3-0.6$ mJy and S/N below 5, and in some cases the optical counterpart is unclear. To be consistent with the LoTSS non-detection, these radio sources must be core-dominated flat-spectrum with radio spectral indices\footnote{The spectral index is defined here as $S_\nu \propto \nu^\alpha$.} of $\alpha>-0.2$, which have been derived using the 3$\sigma$ upper limit on the radio rms measured at 144 MHz. We decided to define our final radio quasar sample in this work solely using the 83 radio detections from LoTSS-DR3. The addition of these extra radio measurements only increases the sample size marginally, and using a single radio frequency (of 144 MHz) ensures a uniformly defined sample. 

\subsection{Rest-frame UV photometry}
\label{subsec:uv_phot}

To obtain rest-frame UV magnitudes of our sample, we used the DESI Legacy Imaging Surveys data release 10 (LS DR10; \citealt{dey2019AJ....157..168D}). This survey provides imaging in three optical bands ($g,r,$ and $z$) and is augmented with four infrared bands from unWISE \citep{Lang2014AJ....147..108L, schlafly2019ApJS..240...30S}. With the reddening corrected photometry from this survey\footnote{\url{https://www.legacysurvey.org/dr10/catalogs/}}, we performed spectral energy distribution (SED) fitting using the EAZY template fitting code \citep{brammer2011ApJ...739...24B} with AGN SED templates from \cite{Brown2019MNRAS.489.3351B}, fixed at the spectroscopic redshift of each quasar. We applied a tophat wavelength window (100 \AA\, width) centered at rest-frame wavelengths of 1450 \AA\, and 4400 \AA\, to extract the UV magnitudes of our quasars. The median UV magnitude values and 16th and 84th percentile errors were obtained through a Monte Carlo approach, where each source is duplicated 500 times with perturbed flux values in each band drawn from the flux errors. We note that due to dust obscuration, the intrinsic UV magnitude values will be brighter than these observed values. 

We performed both a Kolmogorov–Smirnov (KS) test and Anderson-Darling (AD) test on the resulting $M_{1450\text{\AA}}$ distributions of the radio-loud versus radio-quiet samples (see Sect.~\ref{subsubsec:radio_loudness}), as well as radio-detected versus non-detected quasar samples, to investigate their similarity. The KS test is most sensitive to differences near the median of the distribution, while the AD test gives more weight to outliers in the tail. The KS and AD tests showed no statistically significant difference between the $M_{1450\text{\AA}}$ distributions of the radio-loud ($N=26$) and radio-quiet samples ($N=436$; $p_{\text{KS}}=0.76$, $p_{\text{AD}}>0.25$). For the radio-detected ($N=83$) versus non-detected samples ($N=427$), the KS test is border-line significant ($p_{\text{KS}}=0.052$), while the AD test shows a statistically significant difference ($p_{\text{AD}}=0.015$). This difference in $M_{1450\text{\AA}}$ distributions is driven by the sources in the bright and faint end tail with $M_{1450\text{\AA}}<-27$ and $M_{1450\text{\AA}}>-24.5$, including 34 radio non-detected sources and 8 radio detected. Even removing only the 7 radio detected quasars and 11 radio non-detected in the bright tail ($\sim$3\% of our full sample), already increases the $p$-value above the $p<0.05$ significance threshold with $p_{\text{AD}}=0.11$ and $p_{\text{KS}}=0.22$. Since we are dealing with small sample statistics, we decide to consider the full parent sample in this work without removing these sources in the bright M$_{\text{UV}}$ tail. 

\subsection{Low frequency radio properties}
\label{subsec:radio_prop}

\subsubsection{Radio luminosity}
The radio luminosities of our quasars are determined using the standard relation  
\begin{equation}
L_{144\text{MHz}} =  F_{144\text{MHz}} \times \frac{4\pi D_L^2(z)}{(1+z)^{1+\alpha}}     
\end{equation}
with $F_{144\text{MHz}}$ the total flux density, $D_L$ the luminosity distance, and $\alpha$ the radio spectral index (with $(1+z)^{1+\alpha}$ the K-correction). Only a few of our quasars are detected in other radio surveys at higher frequencies (see Sect.~\ref{subsec:other_radio_surveys} for details), and therefore we need to assume a radio spectral index. We adopted the typical radio spectral index of $\alpha=-0.7$ (e.g., \citealt{Hardcastle2016MNRAS.462.1910H}). However, we note that the spectral indices of quasars are known to differ greatly between $-2.0<\alpha<1.0$ (see e.g., \citealt{Gurkan2019A&A...622A..11G, Gloudemans2021A&A...656A.137G, Gloudemans2022A&A...668A..27G}). It is worth emphasizing that flat-spectrum and steep-spectrum radio quasars are physically distinct: the latter exhibit extended radio jets over tens of kpc, while the former arise from pc-scale, young, variable, relativistically beamed jet emission (see reviews by \citealt{Urry1995PASP..107..803U, Odea2021A&ARv..29....3O}).
The impact of this uncertainty in the spectral index is discussed where relevant. This radio-luminosity calculation yielded luminosities of $L_{144\text{MHz}}= 2.5\times10^{25} - 3.3\times10^{28}$ W Hz$^{-1}$ (with a median of $9\times10^{25}$ W Hz$^{-1}$), which if solely generated by star formation would imply incredibly high star formation rates (SFR) of $\gtrsim 500\, M_{\odot}\, \text{yr}^{-1}$, using the relation of \cite{Calistro2017MNRAS.469.3468C} with similar results from \cite{gurkan2018MNRAS.475.3010G, Smith2021A&A...648A...6S, best2023arXiv230505782B, Das2024MNRAS.531..977D}. Recent work on radio AGN from the LOFAR surveys from \cite{best2023arXiv230505782B} and \cite{Yue2024MNRAS.529.3939Y} suggests that at this radio luminosity the radio emission is likely generated by jets and not by star formation. We therefore assume that our radio-detected quasar sample consists of jet-dominated quasars. However, we note that the radio emission can (partly) also be produced by winds and shocks, especially for the lower ($\sim10^{25}$ W Hz$^{-1}$) radio luminosity quasars. The radio sensitivity varies substantially with quasar sky location (see Fig.~\ref{fig:lotss_rms}), and therefore we cannot easily place an overall radio luminosity limit on our sample. The sensitivities of current large sky radio surveys are not sufficient to detect all jet-dominated quasars at this redshift.

\begin{figure*}
    \centering
    \includegraphics[width=1.0\textwidth, trim={0.1cm 0.1cm 0.1cm 0.1cm}, clip]{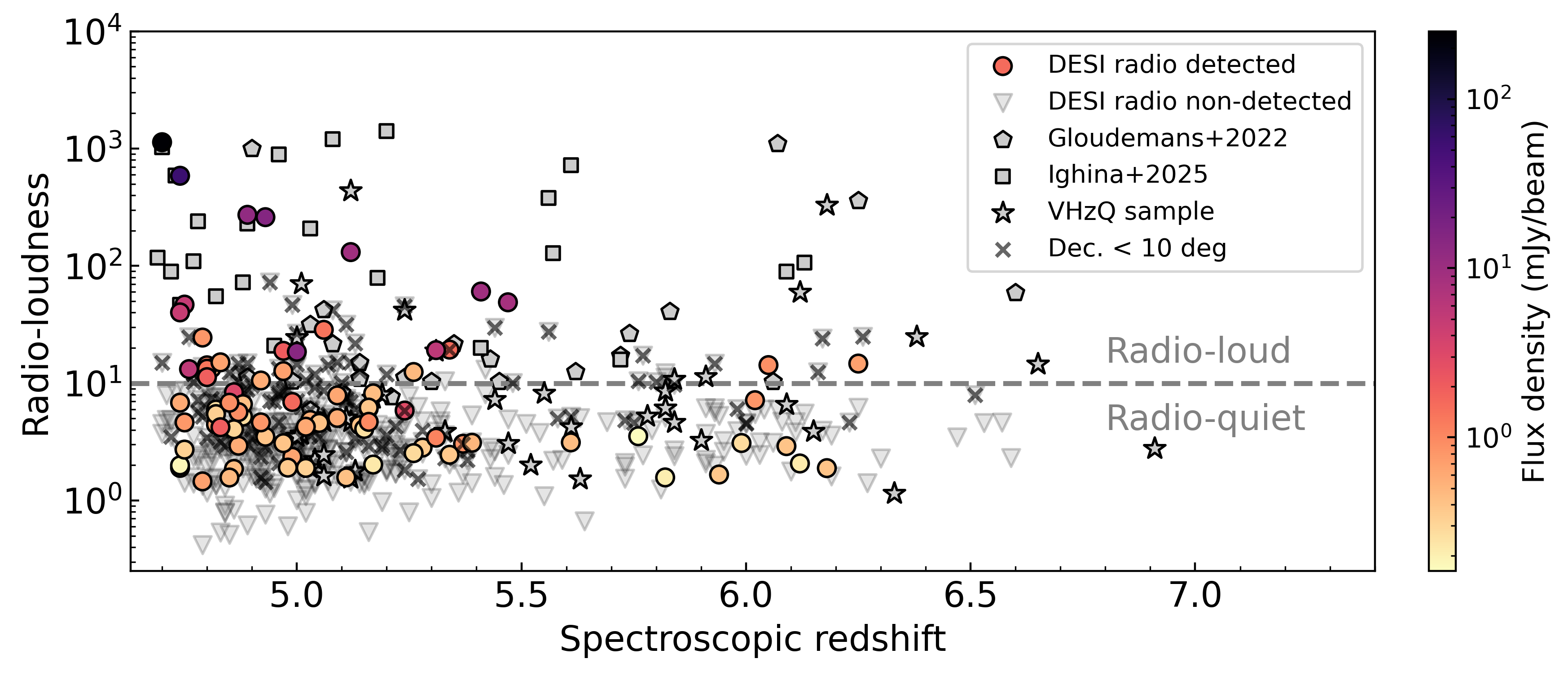}\vspace{-0.0cm}
    \caption{Radio-loudness ($R = F_{5\text{GHz,rest-frame}}/F_{4400\text{\AA}}$) as function of spectroscopic redshift of the DESI high-$z$ quasar sample and literature samples. The radio non-detections are indicated with grey arrows (3$\sigma$ upper limits), and the traditional radio-loudness threshold of $R=10$ is shown by the grey dashed line. This new sample nearly doubles the number of radio quasars at $z>5$. The VHzQ sample is a collection created by \cite{Ross2020MNRAS.494..789R} including all known $z>5$ quasar at the time. Their radio properties are presented in \cite{Gloudemans2021A&A...656A.137G}. The quasars excluded from the ``pure sample'' with a Dec. $<10^{\circ}$ are indicated with a cross.} 
    \label{fig:RL_vs_z}
\end{figure*}

\subsubsection{Radio-loudness}
\label{subsubsec:radio_loudness}

To determine the radio-loudness, we used the definition $R = F_{\nu,5\text{GHz}}/F_{\nu,4400\text{\AA}}$ in rest-frame \citep{Kellermann1989AJ.....98.1195K}, which is generally used in high-$z$ quasar studies (e.g., \citealt{Banados2015ApJ...804..118B, Gloudemans2021A&A...656A.137G, Liu2021ApJ...908..124L}). Other definitions such as $L_{1.4\text{GHz}}/L_{6\mu m}$ are potentially less affected by dust-obscuration of our quasars (e.g., \citealt{Klindt2019MNRAS.488.3109K, Fawcett2020MNRAS.494.4802F}). However, this is difficult to use at high-$z$ due to the lack of reliable photometry at 6 $\mu$m rest-frame. We therefore use the previous definition, which classifies a quasar as radio-loud if $R\geq10$. We again stress that this radio-loudness threshold is somewhat arbitrary and is used in this work mainly to identify the most extreme radio sources in our sample. The radio-loudness of our quasar sample is shown in Fig.~\ref{fig:RL_vs_z}. According to this classical definition, 26 of our radio-detected quasars are classified as radio-loud (5.1\% of the parent sample), while 57 are classified as radio-quiet. The 3$\sigma$ upper limits on the radio-loudness indicate that at least 436 of the 510 quasars (85\% of our sample) are radio-quiet. This places a conservative upper limit of 15\% on the radio-loudness fraction, resulting in a range of $\sim5-15\%$, consistent with previous work (e.g., \citealt{Banados2015ApJ...804..118B, Gloudemans2021A&A...656A.137G}). 

Furthermore, we verified that all but one of the radio-loud sources would have been classified as radio-loud as well according to the $L_{1.4\text{GHz}}/L_{6\mu m} > -4.5$ definition when extrapolating the best fitting SED model as described in Sect.~\ref{subsec:uv_phot}. This sample represents the largest spectroscopic uniform sample of radio quasars at $z\gtrsim5$. A source catalog with the optical and radio measurements of the complete quasar sample will be made public in a forthcoming paper by J. Yang et al. in prep. 

We note that this radio-loudness calculation depends heavily on the assumed spectral index, which differs greatly per quasar. When assuming an ultra steep spectral index ($\alpha=-1.3$), only 10 quasars would be classified as radio-loud, whereas for a flat spectral index ($\alpha=0.0$), the number is as high as 61 quasars. The 26 quasars classified as radio-loud in this work should be considered as a sample of extreme radio bright quasars. Throughout this work, we also compare the radio-detected versus non-detected populations, which provides a complementary, observationally motivated test to examine whether the mere presence of radio emission is correlated with the studied spectral properties.

\subsubsection{Ambiguous radio-loud/radio-quiet classification}
\label{subsubsec:ambiguous_rl_classification}

As apparent in Fig.~\ref{fig:RL_vs_z}, with the radio-loudness definition of $R\geq10$, there are 48 radio non-detected quasars that cannot be securely classified as either radio-loud or radio-quiet with 3$\sigma$ upper limits at $R\geq10$. These ambiguous quasars are mainly located near the equatorial region where the sensitivity of LoTSS decreases (see Fig.~\ref{fig:lotss_rms}). While upper limits in this data set may be treated using survival analysis (e.g., \citealt{Feigelson1985ApJ...293..192F, Schmitt1985ApJ...293..178S, Isobe1986ApJ...306..490I}), they are concentrated at low declination; we therefore instead created a ``pure'' radio-loud and radio-quiet sample that include only quasars at declinations above 10$^{\circ}$, excluding the far majority (39/48) of these ambiguous sources. These ``pure'' samples include 25 radio-loud quasars and 355 radio-quiet quasars, resulting in a slightly more constrained radio-loud fraction of $\sim6-9\%$. 

Furthermore, to take into account these quasars with uncertain radio-loud classification when comparing distributions of the two populations, we generated 10,000 Monte Carlo realizations of the radio-loud/radio-quiet classification by drawing from a half Gaussian distribution centered at zero with a folded (non-negative) tail, scaled by each source's individual uncertainty. The choice of a Gaussian distribution is a simplifying assumption, as the intrinsic radio-loudness distributions of this population is unknown. Especially in the pure sample, most ambiguous sources are close to $R=10$ and therefore have a high chance of being classified as radio-quiet. This classification method yielded a mean radio-loud fraction of 6.1$\pm$1.2\% for the full sample and 6.5$\pm$1.3\% for the pure sample. The uncertainty quoted here is a combination of the classification uncertainty and Poisson counting statistics arising from the small sample sizes. 

\subsubsection{Extended radio sources}
\label{subsec:extended_radio}

From our sample of LoTSS-DR3 detected quasars, the radio images of 6 sources were best fit by multiple Gaussians, giving evidence for resolved radio structures. All of these sources are bright at 144 MHz with total flux densities $\gtrsim 10$ mJy. Their radio images and radio contours superimposed on optical z-band images are shown in Appendix Sect.~\ref{sec:appendix_radio} Figure ~\ref{fig:extended_radio_sources}. We suspect that the second radio component of one of the sources (DESI J134748+440546) is likely related to the galaxy detected in the z-band image, and therefore, this is not an extended radio source. Finally, we note there is one bright 57 mJy radio source (third panel in Fig.~\ref{fig:detected_radio_sources}) that is 1.6\arcsec\, offset from the optical quasar, which could indicate a one-sided lobe structure. Higher resolution VLBI observations are necessary to investigate the radio source structure of these potentially extended radio jets. 

\section{Spectral Analysis}
\label{sec:spec_analysis}

To examine the rest-frame UV spectral properties of this sample, we used the spectra from the DESI Loa data release. These spectra range from 3600 \AA\, to 9824 \AA, covering emission lines such as Ly$\beta$, Ly$\alpha$, Si\textsc{iv}, and \ion{C}{4}. 

\subsection{PyQSOFit Spectral Fitting}
\label{subsec:pyqsofit}

To recover the emission line properties, we performed spectral fitting using the Python QSO fitting code (PyQSOFit v2.0\footnote{\url{https://github.com/legolason/PyQSOFit}}; \citealt{Guo2018ascl.soft09008G, Shen2019ApJ...873...35S}), which was developed specifically for this purpose. The quasar continuum produced by the accretion disk is modeled by a normalized power-law (PL) with a slope of $\beta_{UV}$. Since the wavelength range of DESI only covers a small part of the continuum, we followed the method of \cite{Banados2016ApJS..227...11B} and fixed the PL slope to $\beta_{UV}= -1.5$ and $-1.7$ commonly found for quasars (e.g., \citealt{VandenBerk2001AJ....122..549V}), and took an average between these two results. This method avoids obtaining unphysical values of the continuum slope. Furthermore, PyQSOFit employs an iterative 3$\sigma$ clipping procedure during the continuum fitting, rejecting absorption pixels and thereby minimizing the impact of intervening metal absorption lines on the derived continuum. 

We fitted the important emission lines with one or more Gaussian functions. 
The Ly$\alpha$+\ion{N}{5} complex was fitted with three broad and one narrow line component for Ly$\alpha$ and a single Gaussian for \ion{N}{5}$_{\lambda1240}$. The \ion{C}{4} line was fitted with two broad components, the Si\textsc{iv}$_{\lambda\lambda1397,1402}$ doublet with two Gaussians, and O\textsc{i}$_{\lambda1304}$ with a single Gaussian. The line widths and velocity offsets ($\Delta v_{\mathrm{off}}$) are allowed to vary during the fitting procedure, but are constrained within predefined boundaries. The broad components of Ly$\alpha$ and \ion{C}{4} were constrained by 1400 km s$^{-1} < \mathrm{FWHM} < 14000$ km s$^{-1}$ and $\Delta v_{\mathrm{off}}<10000$ km s$^{-1}$, and the narrow components by 350 km s$^{-1} < \mathrm{FWHM} < 1200$ km s$^{-1}$ and $\Delta v_{\mathrm{off}}<1200$ km s$^{-1}$. Furthermore, the \ion{N}{5}$_{\lambda1240}$, Si\textsc{iv}$_{\lambda\lambda1397,1402}$, O\textsc{i}$_{\lambda1304}$ constraints were slightly tighter with 700 km s$^{-1} < \mathrm{FWHM} < 10000 $ km s$^{-1}$ and $\Delta v_{\mathrm{off}}<7000$ km s$^{-1}$.
Figure \ref{fig:pyqsofit_examples} in Appendix Sect.~\ref{sec:appendix_pyqsofit} shows three example spectra with the best fit from our PyQSOFit routine. The reduced $\chi^2$ values of the Ly$\alpha$ and \ion{C}{4} emission line fits are generally between $0.5-2.0$ with median values of 1.1 and 1.0, respectively.

The \ion{C}{4} line is known to often be shifted to bluer wavelengths in quasars due to outflows (e.g., \citealt{Gaskell1982ApJ...263...79G, Richards2002AJ....124....1R}). To measure the \ion{C}{4} line shift a systemic redshift is needed; however, the brightest line in our spectral range, Ly$\alpha$, is known to be affected by IGM absorption and cannot be used to determine the systemic redshift. Therefore, instead we took the relative offset between Ly$\alpha$ and \ion{C}{4} to estimate the relative \ion{C}{4} velocity shift, noting that these measurements will not take into account the blue-wing asymmetry
commonly exhibited by \ion{C}{4} (e.g., \citealt{Gaskell1982ApJ...263...79G, Wills1993ApJ...415..563W, Richards2011AJ....141..167R}). This offset was determined by taking the peak wavelengths of the best fit of both lines and converting those to rest-frame. The \ion{C}{4} line is shifted outside of the DESI wavelength coverage for $z\gtrsim5.28$, and therefore the velocity offset of \ion{C}{4} can only be determined for the quasars in our sample at $z<5.28$.

Furthermore, we obtained line properties such as rest-frame equivalent width (EW$_{0}$), full-width half maximum (FWHM), and line fluxes of all emission lines. While yielding similar results, we measured the rest-frame EW directly from the spectrum instead of the emission line fit to avoid inaccurate values from suboptimal line fits. The Ly$\alpha$ EW is determined between 1160-1290\AA\ and the \ion{C}{4} EW between 1500-1600\AA\ by integrating over the spectral line divided by the continuum. 

To obtain uncertainties on our spectral properties, we took the same approach as \cite{Shen2019ApJ...873...35S} and perturbed the spectral fluxes at each pixel by randomly drawing from a normal distribution with the standard deviation set by the flux errors. We repeated this process 50 times for each quasar and obtain the 16th, 50th, and 84th percentile values for each spectral property. 

\subsection{Selection of broad absorption line quasars}
\label{subsec:baltools}

To select BAL quasars from our sample, we used the public BAL identification software \textsc{baltools}\footnote{https://github.com/paulmartini/baltools} (see \citealt{Filbert2024MNRAS.532.3669F, Martini2025JCAP...01..137M} for details), which is based on a code originally developed by \cite{Guo2019ApJ...879...72G} for spectroscopic data of the Sloan Digital Sky Survey. The \textsc{baltools} package uses Principal Component Analysis (PCA) to fit the spectrum while iteratively masking any regions with BAL features. The best fit PCA is used to calculate the Balnicity Index (BI; \citealt{Weymann1991ApJ...373...23W}) and Absorption Index (AI; \citealt{Hall2002ApJS..141..267H}), which is used to quantify the strength of the absorption troughs. In this work, we used the AI, which is more sensitive than the BI to narrow absorption troughs (e.g., \citealt{Trump2006ApJS..165....1T}) and is given by
\begin{equation}
    AI = -\int_{25000}^{0} {\Big[}1-\frac{f(\nu)}{0.9}{\Big]} C(\nu)d\nu ,
\end{equation}
with $f(\nu)$ the normalized quasar flux density as a function of the velocity $\nu$ in km s$^{-1}$, $\nu$ the velocity displacement, and $C$ a constant that is 0 unless the trough extends more than 450\,km\,s$^{-1}$, in which case it is set to 1. In our case, the BAL identification could only be applied to quasars at $z\leq5.28$ since the high-ionization \ion{C}{4} line detection is necessary to confidently identify BALs. We classified a quasar as BAL when \ion{C}{4} $AI > 0$, which is satisfied by 140 quasars in our sample (28\%, 140 out of 495 at $z\leq 5.28$). However, only 106 of those are covered by the LoTSS survey. The low-ionization species Mg\textsc{ii}$_{\lambda2800}$ and \ion{Al}{3}$_{\lambda\lambda1854,1862}$, commonly used to identify low-ionization BAL quasars (LoBALs; e.g., \citealt{Weymann1991ApJ...373...23W}), are not covered by our spectra, and therefore we did not separate our BAL quasar sample into LoBAL and high-ionization BAL quasars (HiBALs). We note that the AI calculation also picks up narrower absorption troughs potentially caused by different physical processes, and therefore, the BI calculation is known to be more accurate for BAL identification while being less complete (e.g., \citealt{Knigge2008MNRAS.386.1426K}). We discuss the potential impact of this on our results in Sect.~\ref{subsec:discussion_BALs}. Furthermore, we note that \textsc{baltools} may miss the rare population of extremely high-velocity \ion{C}{4} BALs with $v>25000$ km s$^{-1}$, which become more common among the most luminous quasars, but still represent a small minority of BAL quasars (e.g., \citealt{Bruni2019A&A...630A.111B}).

\section{Results}
\label{sec:results}

The spectral analysis of this high-$z$ quasar sample combined with their radio properties provides the opportunity to investigate the connection between the BLR and radio jets in early quasar systems. In this section, the results are presented separately for the different physical measurements and quasar types. 

\subsection{Lya and \ion{C}{4} EW distributions}
\label{subsec:EW_distributions}

\begin{figure*}
    \centering
    \includegraphics[width=1.0\textwidth, trim={0.1cm 0.1cm 0.1cm 0.1cm}, clip]{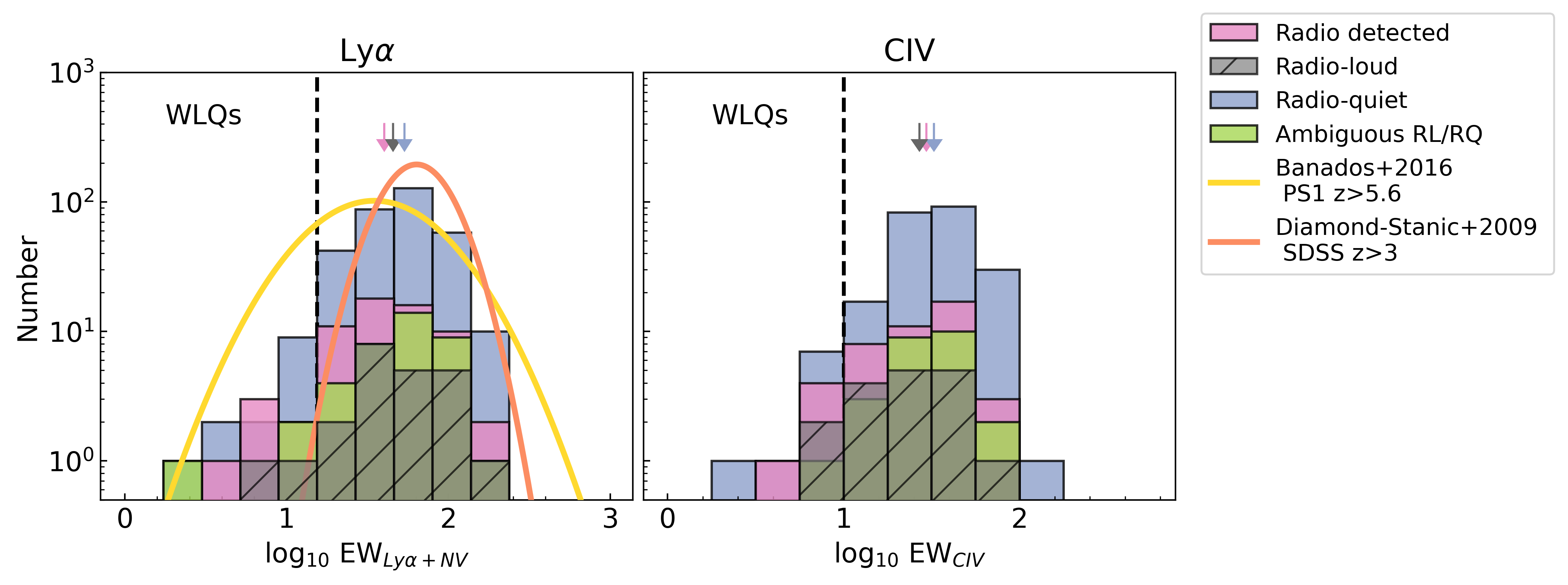}\vspace{-0.0cm}
    \caption{The rest-frame equivalent width distributions of the Ly$\alpha$ and \ion{C}{4} emission lines, highlighting the radio-detected (pink), radio-loud sources (grey hatched), radio-quiet (blue), and ambiguous radio-loud/radio-quiet (green). The median values of the distributions are indicated with the arrows on top. Left: The Ly$\alpha$ WLQ limit of 15.4 \AA\, is indicated by the black dashed line. The yellow and orange lines are high-$z$ quasar EW distributions obtained from previous work of \cite{Banados2016ApJS..227...11B} and \cite{Diamond-Stanic2009ApJ...699..782D}, respectively. These normal distributions are scaled to the same height for comparison. Right: The \ion{C}{4} WLQ limit of 10 \AA\, is again indicated by the black dashed line.}
    \label{fig:EW_hist}
\end{figure*}

The rest-frame Ly$\alpha$+N\textsc{v} and \ion{C}{4} equivalent widths of our quasar sample are shown in Fig.~\ref{fig:EW_hist}. From these distribution we removed the BAL quasars identified by \textsc{baltools} (see Sect.~\ref{subsec:baltools}) and the five quasars with \ion{C}{4} EW measurements below zero. Visual inspection revealed that four of these quasars are likely BAL quasars, and one is heavily affected by telluric lines. 
The radio-detected (pink) and radio-loud (grey hatched) source distributions are compared to the radio-quiet quasar sample (blue). The EW distributions of the ambiguous RL/RQ quasars are indicated in green.
Taking into account the ambiguous RL/RQ quasars by MC sampling (as presented in Sect.~\ref{subsubsec:ambiguous_rl_classification}), the median of the Ly$\alpha$ EW distributions is 46 \AA\, for the radio-loud quasars and 54 \AA\, radio-quiet quasars. This is similar to the previous result found by \cite{Diamond-Stanic2009ApJ...699..782D} of 62 \AA\, and slightly higher than \cite{Banados2016ApJS..227...11B} with an EW distribution peak of 35 \AA. However, these measurements of Ly$\alpha$ EW cannot be directly compared, since these quasar samples have different redshift distributions and are therefore differently affected by IGM absorption. 

We performed both KS and AD tests to compare the EW distributions across three sample pairs: radio-loud versus radio-quiet, the ``pure'' subsamples restricted to Dec.$>10^{\circ}$, and the radio-detected versus non-detected quasars. The KS test is most sensitive to differences near the median of the distribution, while the AD test gives more weight to outliers in the tail. For the radio-loud and radio-quiet samples, we performed these tests 10,000 times using the different MC realizations and report the median $p$-value. 
Because the radio-loud/radio-quiet classification is probabilistic under our MC scheme, we report the mean number of sources assigned to each class across MC realizations, denoted $N_{\rm mean}$.

Both the KS and AD tests found a significant difference ($p<0.05$) between the radio-detected and radio non-detected EW distributions, for both Ly$\alpha$ ($p_{\text{KS}}=0.016$; $p_{\text{AD}}=0.014$ with $N=$ 63 and 341) and \ion{C}{4} ($p_{\text{KS}}=0.041$; $p_{\text{AD}}=0.010$ with $N=$44 and 231). In contrast, we did not find a similar significant difference when comparing the radio-loud and radio-quiet samples. The KS and AD tests for the Ly$\alpha$ EW distributions resulted in $p>0.05$ for both the radio-loud versus radio-quiet samples ($p_{\text{KS}}=$0.41; $p_{\text{AD}}>0.25$  with $N_{\rm mean}=$27 and 377) and the pure radio-loud versus radio-quiet samples ($p_{\text{KS}}=0.30$; $p_{\text{AD}}>0.25$ with $N_{\rm mean}=$22 and 290). The AD tests on the \ion{C}{4} EW distributions of those samples resulted in $p_{\text{AD}}=0.058$ for the normal sample and $p_{\text{AD}}=0.040$ for the pure sample, indicating these distributions could be significantly different. However, this was not confirmed by the KS test with $p_{\text{KS}}\sim0.12$ for both.
These results suggest a potential difference in the Ly$\alpha$ and \ion{C}{4} EW distributions between radio-detected and non-detected quasars, with radio-detected quasars exhibiting, on average, smaller EWs. A similar trend may also be present for the radio-loud versus radio-quiet comparison, although a larger sample size would be required to confirm this. 

\begin{figure*}
    \centering
    \includegraphics[width=1.0\textwidth, trim={0.1cm 0.1cm 0.1cm 0.1cm}, clip]{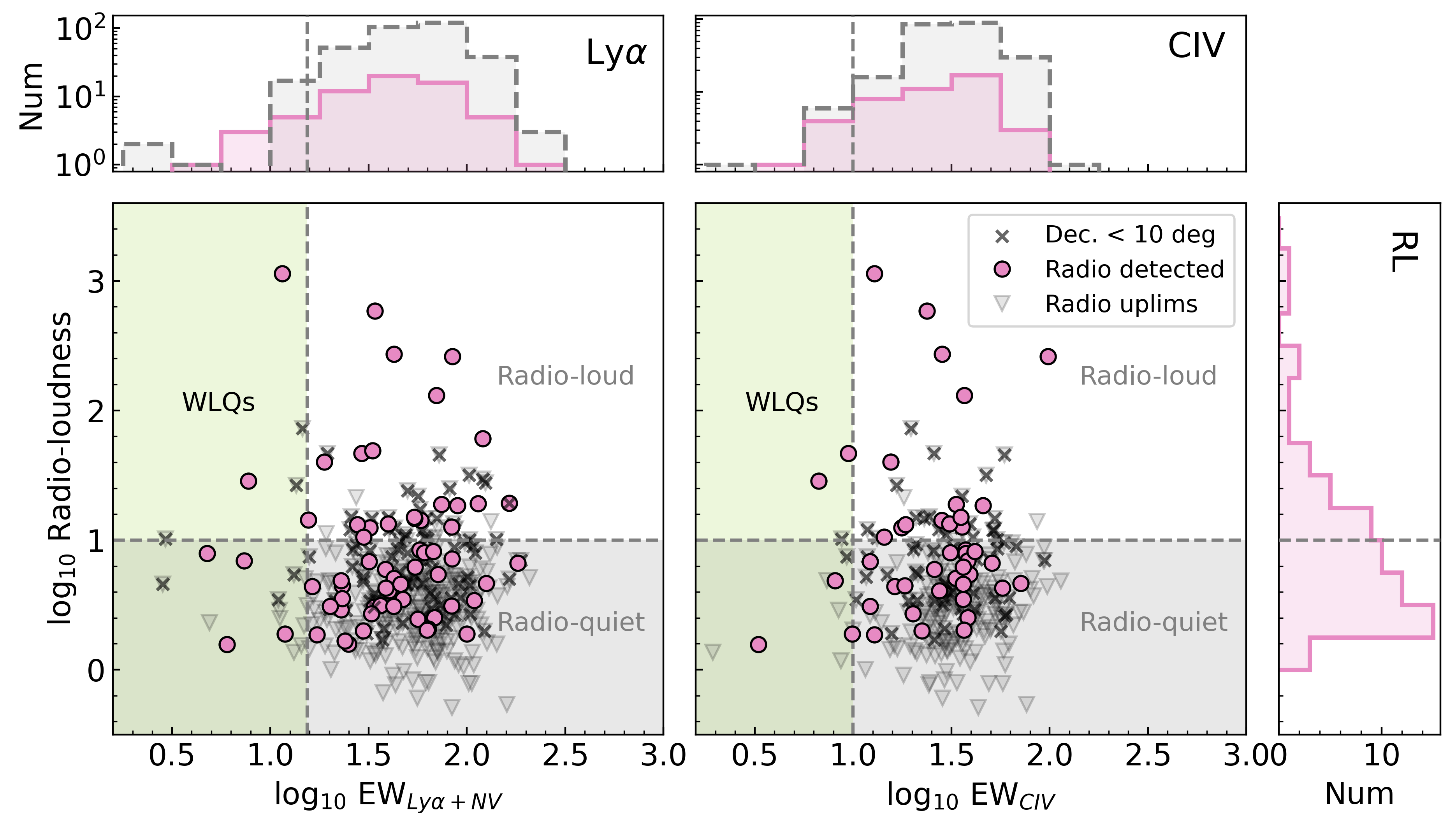}\vspace{-0.0cm}
    \caption{Radio-loudness of our DESI quasar sample versus Ly$\alpha$ EW (left) and \ion{C}{4} EW (right). The radio non-detected quasars are indicated with 3$\sigma$ upper limits, and the crosses indicate the quasars below $10^{\circ}$ declination where the LOFAR sensitivity drops significantly. The top panels show the EW distributions of the radio detected versus radio non-detected population, while the right most panel shows the radio-loudness distribution of the radio detected quasars. This figure does not include the identified BAL quasars, since their EW measurements are uncertain.} 
    \label{fig:RL_vs_EW}
\end{figure*}

The general distributions of radio-loudness versus Ly$\alpha$ and \ion{C}{4} EW are shown in Fig.~\ref{fig:RL_vs_EW}. We did not find a significant correlation between the radio-loudness values of the radio-detected quasars and their Ly$\alpha$ and \ion{C}{4} EW, with a Spearman's correlation coefficient of 0.17 and $p$-value of 0.18 for Ly$\alpha$ and a coefficient of 0.07 and $p$-value of 0.66 for \ion{C}{4}. 

Early comparisons between low-$z$ radio-loud and radio-quiet quasars found mixed evidence for systematic spectral differences in the rest-frame UV. For example, \cite{Steidel1991ApJ...382..433S} and \cite{Corbin1992ApJ...391..577C} found no strong differences in the overall UV spectral properties of radio-loud and radio-quiet quasars, whereas \cite{Francis1993AJ....106..417F} reported (modestly) higher Ly$\alpha$ and \ion{C}{4} EWs in radio-loud quasars.
More recent work, however, has shown that UV emission-line properties of $z\lesssim3$ quasars are more closely linked to the accretion state of the quasar than to radio loudness itself. This accretion state sequence was originally identified via the optical Eigenvector 1 correlation space \citep{Boroson1992ApJS...80..109B}, later extended into the broader ``quasar main sequence'' framework \citep{Sulentic2000ApJ...536L...5S} and connected to UV C~\textsc{iv} properties \citep{Wills1999ApJ...515L..53W, Shen2014Natur.513..210S, Sulentic2017A&A...608A.122S}.
\cite{Richards2011AJ....141..167R}, \cite{Kratzer2015AJ....149...61K}, and \cite{Rankine2021MNRAS.502.4154R} found significant differences in the C~\textsc{iv} blueshift and EWs of RL and RQ quasars, with RL quasars being biased to lower C~\textsc{iv} blueshifts and larger C~\textsc{iv} EWs, although to a much lesser extent.
However, \cite{Rankine2021MNRAS.502.4154R} highlighted that radio detected quasars can be found everywhere in \ion{C}{4} parameter space, hindering the prospect of identifying an individual radio-loud quasar based on its UV emission line properties.
The C~\textsc{iv} velocity shifts of our sample will be discussed in Sect.~\ref{subsec:results_civ_shift}, and the apparent increased fraction of weak-line quasars amongst the radio quasar population in Sect.~\ref{subsec:results_WLQs}, which is (at least partly) driving the observed statistical difference in EW distributions.

\subsection{Radio emission of WLQs}
\label{subsec:results_WLQs}

\begin{figure*}
    \centering
    \includegraphics[width=0.8\textwidth, trim={0.1cm 0.1cm 0.1cm 0.1cm}, clip]{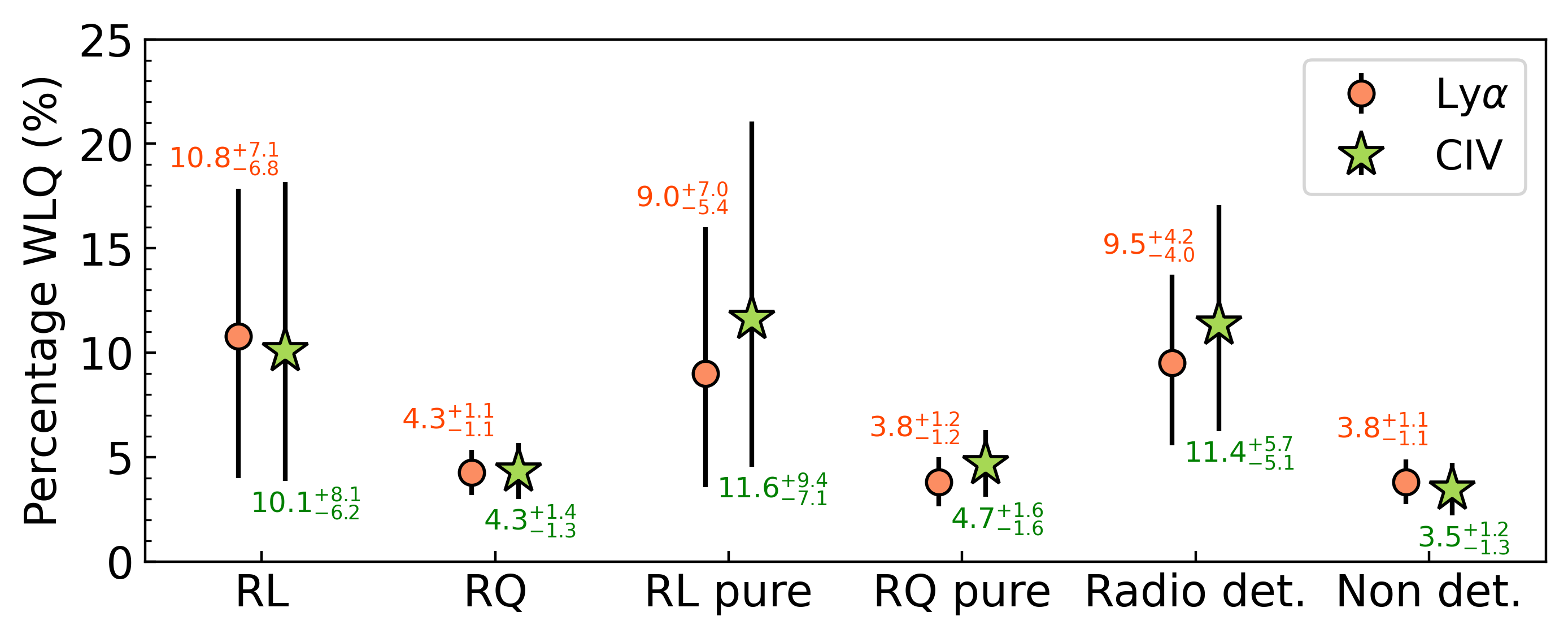}\vspace{-0.0cm}
    \caption{Percentages of weak-line quasars in different subsamples, excluding the identified BAL quasars. The radio-loud quasars are defined as $R = F_{\nu,5\text{GHz}}/F_{\nu,4400\text{\AA}} \geq 10$ whereas the radio-quiet quasars have either $R$ or $R_{\text{lim},3\sigma}<10$. The ``pure'' samples contain only quasars with Dec. $> 10^{\circ}$. The errors have been estimated from random sampling Poisson distributions, and reflect the potential scatter due to small sample size. In the case of the RL and RQ subsamples, the percentages and errors also take into account the ambiguous radio non-detected sources above the radio-loudness limit.}
    \label{fig:WLQ_percentages}
\end{figure*}

To identify the WLQs in our sample, we followed the definition of \cite{Diamond-Stanic2009ApJ...699..782D} with a limit of 15.4\,\AA\ for the Ly$\alpha$ EW and 10\,\AA\ for the \ion{C}{4} EW. This Ly$\alpha$ EW also includes varying contributions from \ion{N}{5}$_{\lambda1240}$ and \ion{Si}{2}$_{\lambda1263}$ (e.g., \citealt{Leighly2007ApJS..173....1L}). 
These limits yielded 19 WLQs from the Ly$\alpha$ measurements and 13 from the \ion{C}{4} measurements within the LOFAR footprint. Out of these, 6 quasars at $z\leq5.28$ satisfied both Ly$\alpha$ and \ion{C}{4} criteria. An example spectrum of one of these quasars, DESI J002633.61+251653.1, is shown in the bottom panel of Fig.~\ref{fig:example_BAL_WLQ} together with its radio contours. We note that the limits from \cite{Diamond-Stanic2009ApJ...699..782D} have been determined at lower redshift ($3<z<5$) where the Ly$\alpha$ line is less affected by IGM absorption. The \ion{C}{4} EW measurements are therefore more reliable, however, they are only available at $z\leq5.28$ due to spectral coverage. 

To investigate any connection between the WLQ phenomenon and radio emission, we determined the fractions of WLQs for our different subsamples. The results are shown in Fig.~\ref{fig:WLQ_percentages}. The reported WLQ fractions of the radio-loud and radio-quiet samples are the mean of the derived distributions from the MC classification method as described in Sect.~\ref{subsubsec:ambiguous_rl_classification}. The uncertainties on the fractions reflect both the classification uncertainty and Poisson counting statistics, propagated jointly via Monte Carlo resampling to yield asymmetric 68\% confidence intervals.

While at face value the fractions of WLQs in the radio-detected and radio-loud quasar samples are $\sim2\times$ higher than the radio non-detected and radio-quiet quasar samples, the large errors indicate these differences are only $\sim1\sigma$ significant. Therefore, this difference could be due to the small sample size. The most significant difference is found comparing the radio detected WLQ fraction ($9.5^{+4.2}_{-4.0}$\% for Ly$\alpha$ and $11.4^{+5.7}_{-5.1}$\% for \ion{C}{4}) with the radio non-detected WLQ fraction ($3.8^{+1.1}_{-1.1}$\% for Ly$\alpha$ and $3.5^{+1.2}_{-1.3}$ for \ion{C}{4}), yielding a significance of 1.3$\sigma$ for Ly$\alpha$ and 1.4$\sigma$ for \ion{C}{4}.

The percentages of WLQs found in this work are generally lower than previously reported in high-$z$ quasar studies (e.g., \citealt{Banados2014AJ....148...14B, Banados2016ApJS..227...11B, Gloudemans2022A&A...668A..27G}), ranging from 14-38\%. Our fractions are more similar to \cite{Diamond-Stanic2009ApJ...699..782D}, who studied a sample of 74 SDSS selected WLQs at $3 \leq z \leq 5$ and found their WLQ fraction increases from 1.3\% at $3 < z< 4.3$ to 6.2\% at $4.2 < z < 5.0$. The implications of our findings are further discussed in Sect.~\ref{sec:discussion}.

\subsection{Radio emission of BAL QSOs}
\label{subsec:results_BALs}

\begin{figure*}
    \centering
    \includegraphics[width=1.0\textwidth, trim={0.1cm 0.2cm 0.1cm 0.1cm}, clip]{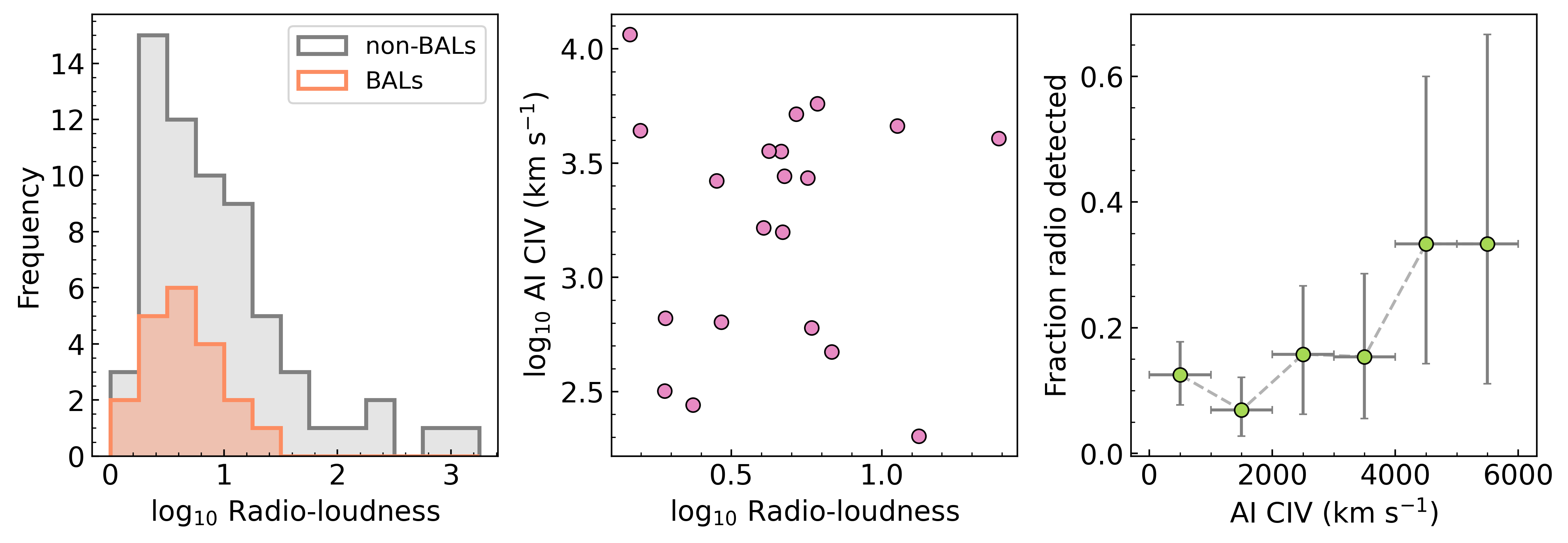}\vspace{-0.0cm}
    \caption{Left: Radio-loudness distribution of quasars classified as BAL and no BAL. While there are a few BAL quasars above the traditional radio-loudness limit, they do not populate the tail of the radio-loudness distribution. Middle: The radio-loudness of our BAL quasars versus their \ion{C}{4} absorption index. There is no significant correlation. Right: The fraction of radio detected BAL quasars as a function of the \ion{C}{4} absorption index, showing a tentative (non-significant) positive correlation. The errors have been estimated using Monte Carlo simulations drawn from Poisson distributions. } 
    \label{fig:RL_BAL_QSO_combined}
\end{figure*}

As described in Sect.~\ref{subsec:baltools}, we identified 140 BAL quasars in our $4.80 \leq z \leq 5.28$ sample, 106 of which are covered by LoTSS. The fraction of BAL quasars detected at 144 MHz is 18.9\% (20/106), which is similar to the radio detected fraction of non-BAL quasars of 15.5\% (44/284) at $z\leq5.28$.
The left panel of Figure \ref{fig:RL_BAL_QSO_combined} shows the radio-loudness distribution of BAL versus no BAL quasars, which demonstrates that BAL quasars are predominantly radio-quiet with no sources at $R \geq 25$. 
This is also reflected in the fraction of BAL quasars in the different subsamples. In the radio-loud sample, the fraction of BAL quasars is 16.3$\pm$10.1\% (15.0$\pm$10.0\% for the pure sample), while in the radio-quiet sample, it is 27.9$\pm$3.1\% (26.8$\pm$3.5\% for the pure sample). However, this observed difference is only $\sim1\sigma$ significant due to low number statistics, and this is not apparent in the radio-detected sample with a fraction of 31$\pm$8\% (20/64) BAL quasars compared to 26$\pm$3\% (86/326) for the radio non-detected quasar sample.

\begin{figure*}
    \centering
    \includegraphics[width=0.85\textwidth, trim={0.1cm 0.2cm 0.1cm 0.1cm}, clip]{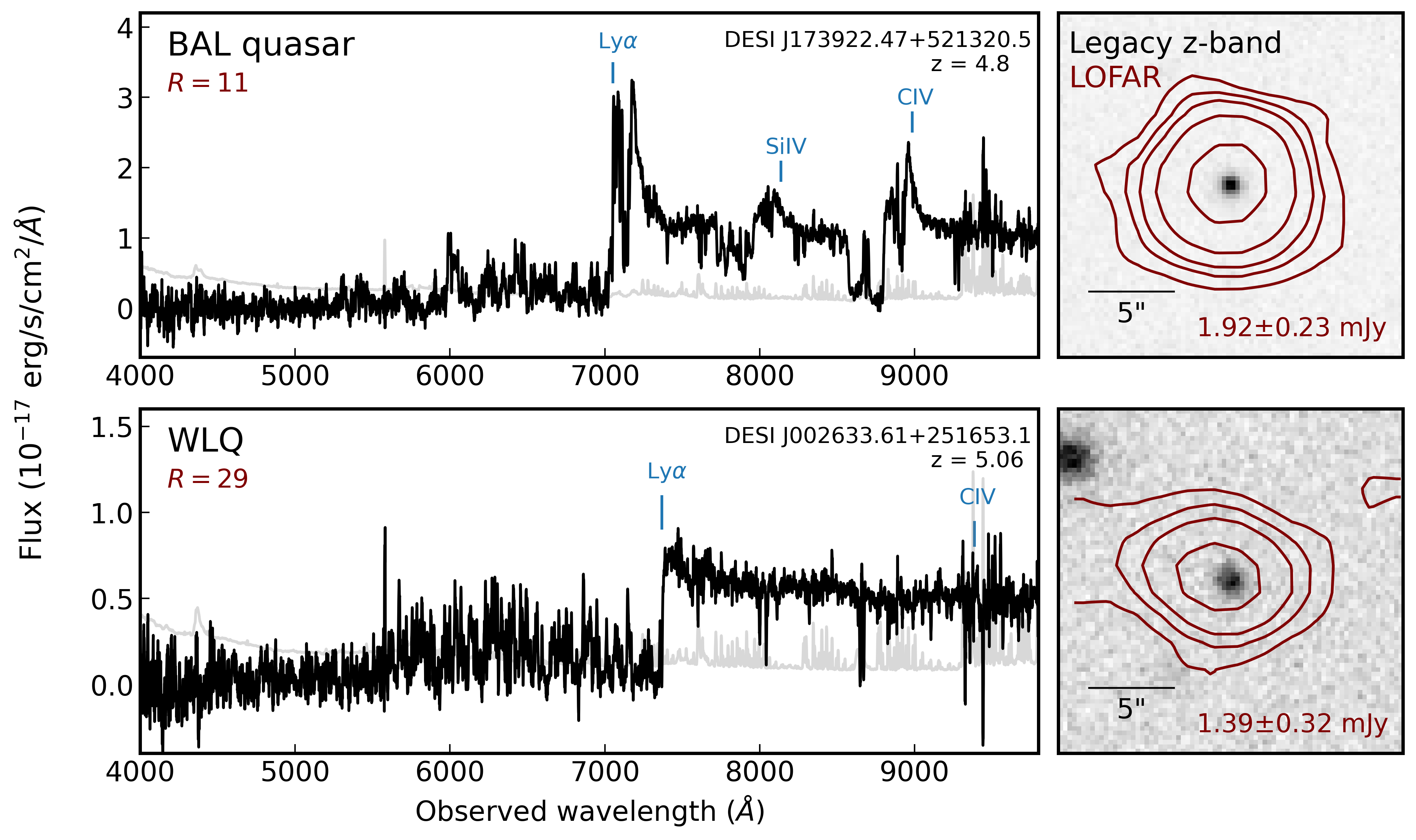}\vspace{-0.0cm}
    \caption{Example spectra of a radio-loud BAL quasar (top) and WLQ (bottom), illustrating the range of optical spectral properties within the radio-loud quasar population. The grey line shows the error on the spectrum. The right panels show the optical z-band images of the quasars with radio contours from LOFAR. For each image, the radio contours are drawn at [1, 2, 3, 5, 10]$\times$ the radio rms, and their total flux density is given in the bottom right corner.} 
    \label{fig:example_BAL_WLQ}
\end{figure*}

An example spectrum and radio contours of one of the radio-loud BAL quasars (DESI J173922.47+521320.5 from \citealt{Yang2023ApJS..269...27Y}) are shown in the top panels of Fig.~\ref{fig:example_BAL_WLQ}. As shown in the middle panel of Fig.~\ref{fig:RL_BAL_QSO_combined}, we did not find any correlation between the absorption index AI and radio-loudness (including 20 quasars), with a Spearman's correlation coefficient of 0.02 and a $p$-value of 0.92 when using the radio-loudness measurements from the full sample. Repeating this calculation for the pure sample yielded similar results with a correlation coefficient of 0.07 and a $p$-value of 0.76. However, we did find a (statistically insignificant) increase in the radio detected fraction with increasing absorption index, as shown in the right panel of Fig.~\ref{fig:RL_BAL_QSO_combined}. We required at least 5 sources per bin and therefore only measured the radio fraction up to 6000 km s$^{-1}$. Since the error bars are large, we determined the Spearman's correlation statistic and $p$-value by sampling the radio detected fraction 10,000 times from a normal distribution set by the uncertainty. This resulted in a correlation coefficient of 0.50$\pm$0.38 and a $p$-value of 0.29$\pm$0.28. The apparent increase in radio detected fraction is therefore statistically insignificant with the current sample size. 

Generally, these results are in line with previous work, which indicates that BAL quasars are predominantly radio-quiet, however, some of them are radio-loud (e.g., \citealt{Brotherton1998ApJ...505L...7B, Becker2000ApJ...538...72B, Morabito2019A&A...622A..15M}). There is no general consensus yet on whether orientation or evolution is causing the differences between BAL and non-BAL quasars. In line with our results, previous work from \cite{Morabito2019A&A...622A..15M} found that the radio detected fraction increases with balnicity index (BI), while not finding any correlation between BI and radio-power or radio-loudness. They concluded that the radio emission and absorption are distinct phenomena, but the underlying process that creates them might be linked. This study suggests this might be the case as well at high redshift. Subsequent work by \cite{Petley2022MNRAS.515.5159P, Petley2024MNRAS.529.1995P} suggests that instead the enhanced radio detection may be due to radio emission related to AGN wind shocks in the ISM. However, the radio luminosities of the quasars in this work (especially at $L_{150\text{MHz}} > 10^{26}$ W Hz$^{-1}$) are likely too high for AGN winds to be the only source of radio emission, and therefore must contain jets.

\subsection{\ion{C}{4} blueshift of radio quasars}
\label{subsec:results_civ_shift}

\begin{figure}
    \centering
    \includegraphics[width=1.0\columnwidth, trim={0.1cm 0.1cm 0.1cm 0.1cm}, clip]{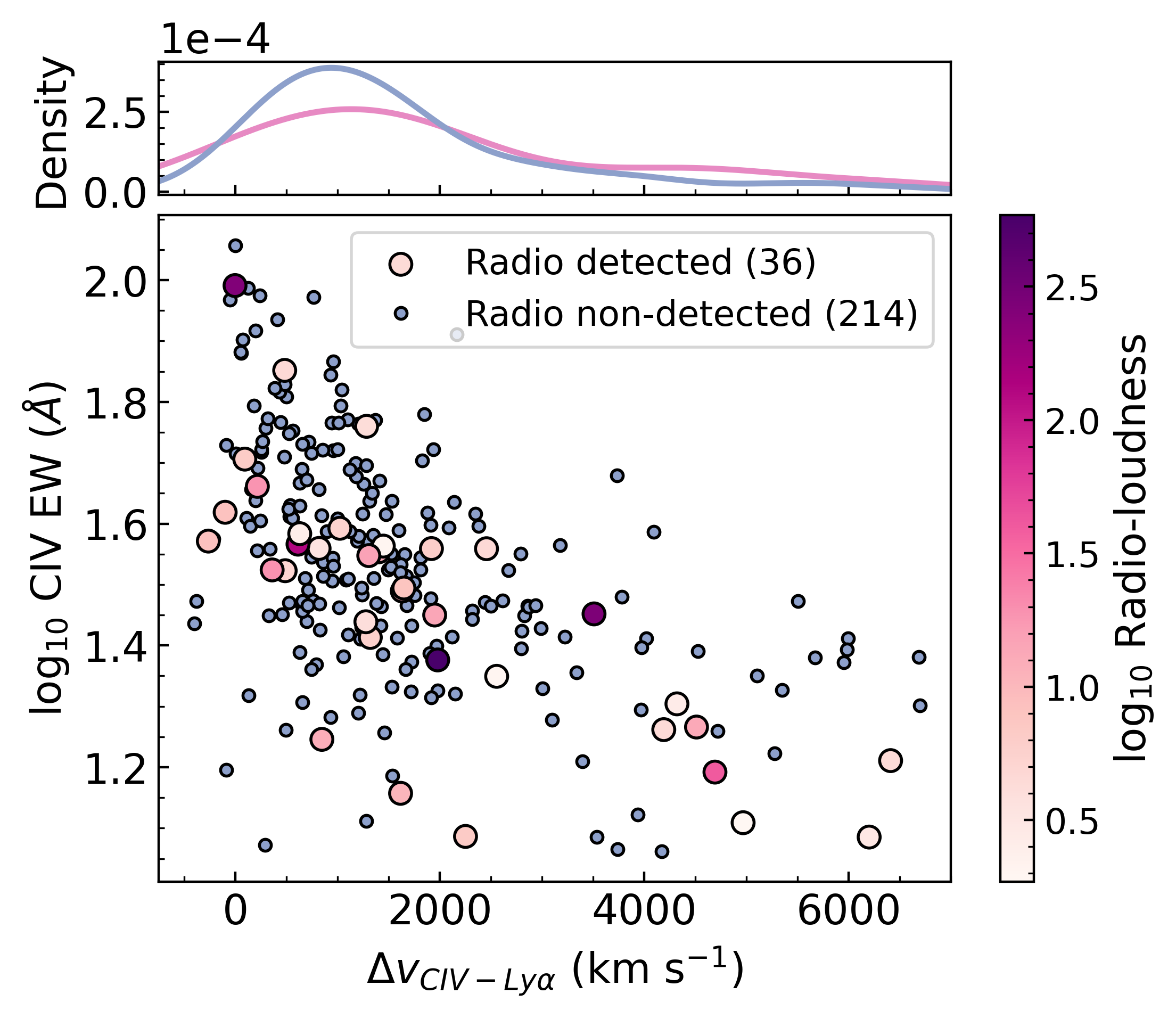}\vspace{-0.0cm}
    \caption{\ion{C}{4} rest-frame EW versus the velocity shift between the Ly$\alpha$ and \ion{C}{4} emission line for radio detected (pink) and non-detected (blue) quasars. The radio quasars cover a similar parameter space as the radio non-detected quasars. The colorbar indicates the radio-loudness of the radio detected quasars. The top panel shows the kernel density distributions of $\Delta v_{\text{C\mbox{\textsc{iv}}}-\text{Ly}\alpha}$ of the radio detected and non-detected samples.}
    \label{fig:CIV_blueshift}
\end{figure}

To examine any correlation between \ion{C}{4} blueshift (indicative of outflow strength) and radio emission of our quasar sample, we used the velocity offsets between Ly$\alpha$ and \ion{C}{4} as determined from spectral fitting in Sect.~\ref{subsec:pyqsofit}. To avoid inaccurate measurements, we removed sources from the sample that have i) a \ion{C}{4} EW below 10 \AA\,, ii) a Ly$\alpha$ EW below 15.4 \AA, and iii) are identified as BAL quasar. The distributions of radio detected and non-detected quasars in \ion{C}{4} EW and velocity shift space are shown in Fig.~\ref{fig:CIV_blueshift} with their radio-loudness color-coded. From this figure, it is apparent that the radio quasars cover a similar parameter space as the radio non-detected quasar sample, and the radio-loudest quasars are spread out over the full parameter space. 
To compare the distributions, we again applied the KS and AD tests to the C~\textsc{iv} velocity shift across our three sample pairs. None of the three comparisons yield a statistically significant difference: radio-loud ($N_{\text{mean}}=16$) versus radio-quiet ($N_{\text{mean}}=234$, $p_{\text{KS}}=0.54$, $p_{\text{AD}}>0.25$), the pure subsamples ($N_{\text{mean}}=14$ versus 179, $p_{\text{KS}}=0.70$, $p_{\text{AD}}>0.25$), and radio-detected ($N=36$) versus non-detected ($N=214$, $p_{\text{KS}}=0.26$, $p_{\text{AD}}>0.25$).
In addition, we performed a two-dimensional KS test using the public code \textsc{ndtest}\footnote{Written by Zhaozhou Li, \url{https://github.com/syrte/ndtest}, based on \cite{Peacock1983MNRAS.202..615P, Fasano1987MNRAS.225..155F, Press2007nras.book.....P}.} using both the \ion{C}{4} velocity shift and the \ion{C}{4} EW, again finding no significant differences for any of the samples with $p>0.15$ in all cases.

We therefore did not find that these radio-loud quasars are skewed towards lower \ion{C}{4} blueshifts as found by, for example \cite{Rankine2021MNRAS.502.4154R} and \cite{Jackson2026MNRAS.546ag065J}; however, they also note that radio sources can be found almost everywhere in \ion{C}{4} emission space. We caution that our \ion{C}{4} blueshift measurements (peak-to-peak offset relative to Ly$\alpha$) do not take into account the extended blue wing of \ion{C}{4}, unlike these literature works that account for this using systemic redshifts. This difference in methodology may influence the results.
Furthermore, previous works report both a decrease in radio detected fraction with \ion{C}{4} blueshifts (e.g., \citealt{Marziani1996ApJS..104...37M, Richards2002AJ....124....1R}), and an increase (e.g., \citealt{Rankine2021MNRAS.502.4154R}). However, we found a fairly constant radio-detected fraction of $10-30\%$ across all \ion{C}{4} blueshifts, which is on the higher end of the range of values identified in \cite{Richards2002AJ....124....1R}. The \ion{C}{4} blueshift is known to correlate with the bolometric luminosity (e.g., \citealt{Rankine2020MNRAS.492.4553R}), which we found as well with a correlation coefficient of 0.12, and $p$-value of 0.03 for our full sample of 316 quasars (including 250 within the LoTSS footprint). We caution that the low-number statistics impose strong limitations on our analysis and may hinder our ability to identify the statistical differences between radio-quiet and radio-loud quasars that have been reported at low redshift.

\subsection{Radio-loudness evolution}
\label{subsec:RL_evolution}

To examine any radio-loudness evolution, we built a lower-$z$ quasar sample at $2\leq z \leq 2.5$ to compare to our $z\sim5$ sample. We picked this redshift range because it coincides with the peak in AGN space density and offers a sufficiently large sample for statistical analysis. We selected these lower-$z$ quasars from the public DESI DR1 value-added catalog (VAC\footnote{\url{https://data.desi.lbl.gov/doc/releases/dr1/vac/agnqso/}}, S. Juneau et al., in prep.), which includes quasars from all surveys and classifications and redshift estimates from the DESI Redrock pipeline (Bailey et al., in prep.) refined using QuasarNet and the Mg\textsc{ii} post processing pipeline \citep{Chaussidon2023ApJ...944..107C, Alexander2023AJ....165..124A}. 

\begin{figure*}
    \centering
    \includegraphics[width=1.0\textwidth, trim={0.1cm 0.1cm 0.1cm 0.1cm}, clip]{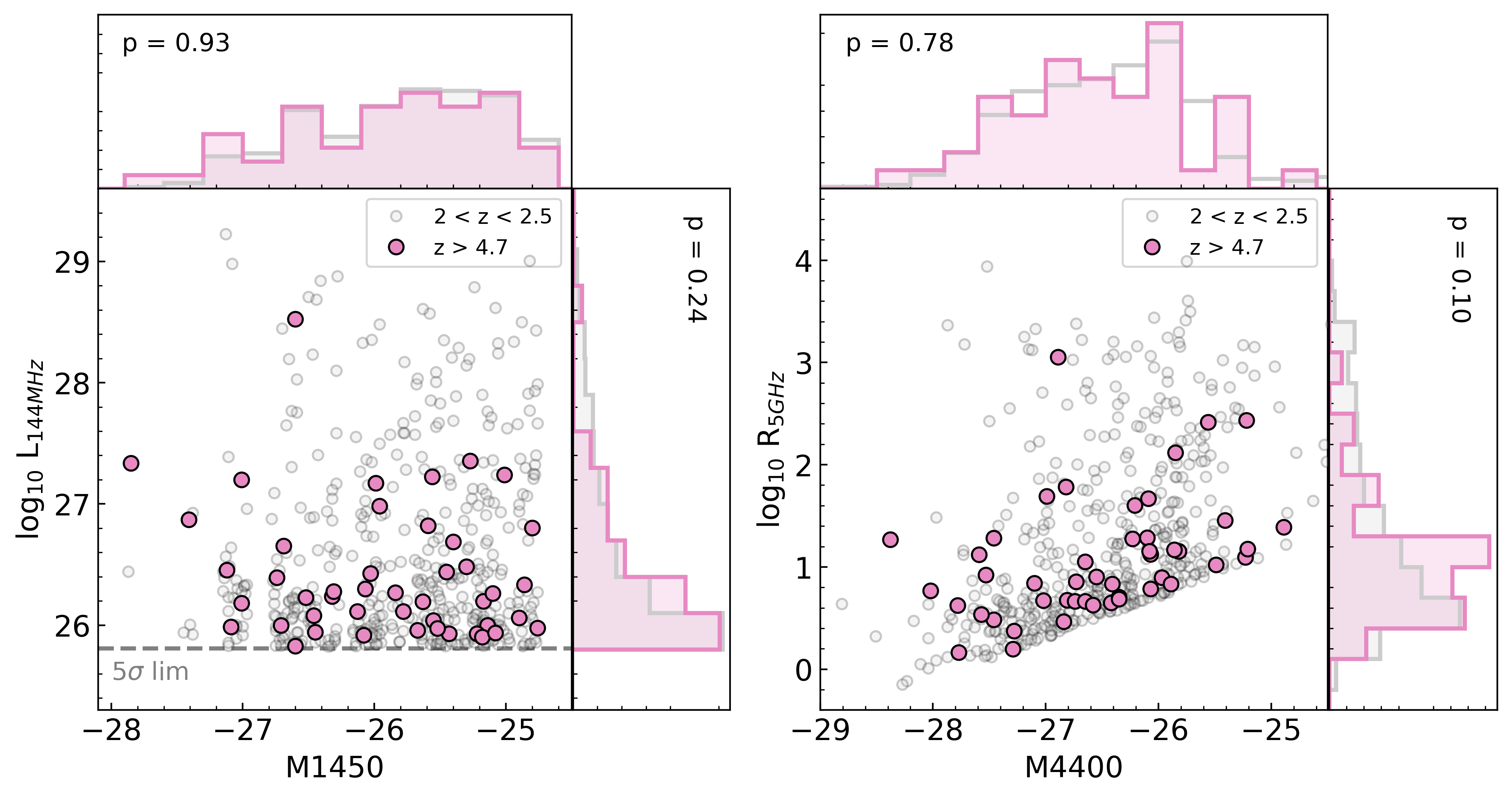}\vspace{-0.0cm}
    \caption{Comparison of the radio properties of our $z>4.7$ quasar sample and a rest-frame UV luminosity matched $2\leq z\leq 2.5$ sample. Left: Radio-luminosity versus M$_{1450\text{\AA}}$ for the low- and high-$z$ quasar sample. The top and side panel show the normalized distributions of both samples and the $p$-value obtained from a two-sided KS test. The distributions are statistically similar, however, the high-$z$ sample does not show the same tail out to high radio-luminosities. Right: Radio-loudness versus M$_{4400\text{\AA}}$ for the low- and high-$z$ samples. Again, the distributions are statistically similar and do not show any evidence for a radio-loudness evolution.} 
    \label{fig:RL_evolution}
\end{figure*}

We cross-matched all sources classified as a quasar (i.e. ``SPECTYPE = QSO'') between $2\leq z \leq 2.5$ with the LoTSS-DR3 catalog within 1.5\arcsec, which yields radio detections for 13654 quasars (out of a parent sample of 324683 quasars, i.e. $\sim$4\% LoTSS detection rate). To mimic the radio detection threshold at $z\sim5$, we applied a radio luminosity cut of $L_{150\text{MHz}} > 6.5\times10^{25}$ W Hz$^{-1}$, which is 5$\sigma$ above the median rms sensitivity. This resulted in a lower-$z$ sample of 3785 quasars. We performed the same SED fitting technique as described in Sect.~\ref{subsec:uv_phot} using the LS DR10 photometry to obtain their rest-frame UV magnitudes. For consistency, we included only the 47 LoTSS-DR3 detected quasars above $L_{150\text{MHz}} > 6.5\times10^{25}$ W Hz$^{-1}$ in the high-$z$ sample. Finally, to ensure the rest-frame UV magnitude distributions of the low- and high-$z$ sample are statistically similar, we randomly sampled 10 lower-$z$ quasars with an M$_{1450\text{\AA}}$ around $\pm0.05$ of each quasar in the high-$z$ sample. This process created the final lower-$z$ quasar sample of 434 sources (not including any duplicates). We determined their radio luminosities and radio-loudness as described in Sect.~\ref{subsec:radio_prop}. The results are shown in Fig.~\ref{fig:RL_evolution}. This figure demonstrates that both the radio luminosity and radio loudness distributions do not show evidence of any evolution. A two-sided KS test (including 47 quasars at $z\sim5$ and 434 quasars at $z\sim2$) indicated that both of these could have been drawn from the same underlying distributions with $p=0.24$ for the radio-luminosity and $p=0.10$ for the radio-loudness. Notably, the high-$z$ sample does not show the same tail out to high radio luminosity and loudness as the lower-$z$ sample, however, that could be due to the relatively small number of radio sources in the high-$z$ sample. We also performed the AD test, which resulted in $p$-values of 0.15 and 0.09 for the radio-luminosity and radio-loudness, respectively, which are again not significant.

Finding no evidence for radio loudness evolution is in line with most previous works (e.g., \citealt{Banados2015ApJ...804..118B, Gloudemans2021A&A...656A.137G, Liu2021ApJ...908..124L}) and establishes the fact that these radio sources are already active within $\sim$1 Gyr after the Big Bang. This result is perhaps unsurprising, as previous studies have suggested that high-$z$ quasars generally exhibit UV and optical properties similar to those of lower-$z$ quasars, both in their emission-line strengths and continuum shapes (e.g., \citealt{Pentericci2003A&A...410...75P, Fan2004AJ....128..515F, Shen2019ApJ...873...35S}). Probing fainter radio sources with future radio telescopes such as the Square Kilometer Array (SKA; \citealt{Dewdney2009IEEEP..97.1482D}) is necessary for more detailed studies of the radio-loudness evolution.

\subsection{Spectral stacking}
\label{subsec:stacks}

\begin{figure*}
    \centering
    \includegraphics[width=1.0\textwidth, trim={0.1cm 0.1cm 0.1cm 0.1cm}, clip]{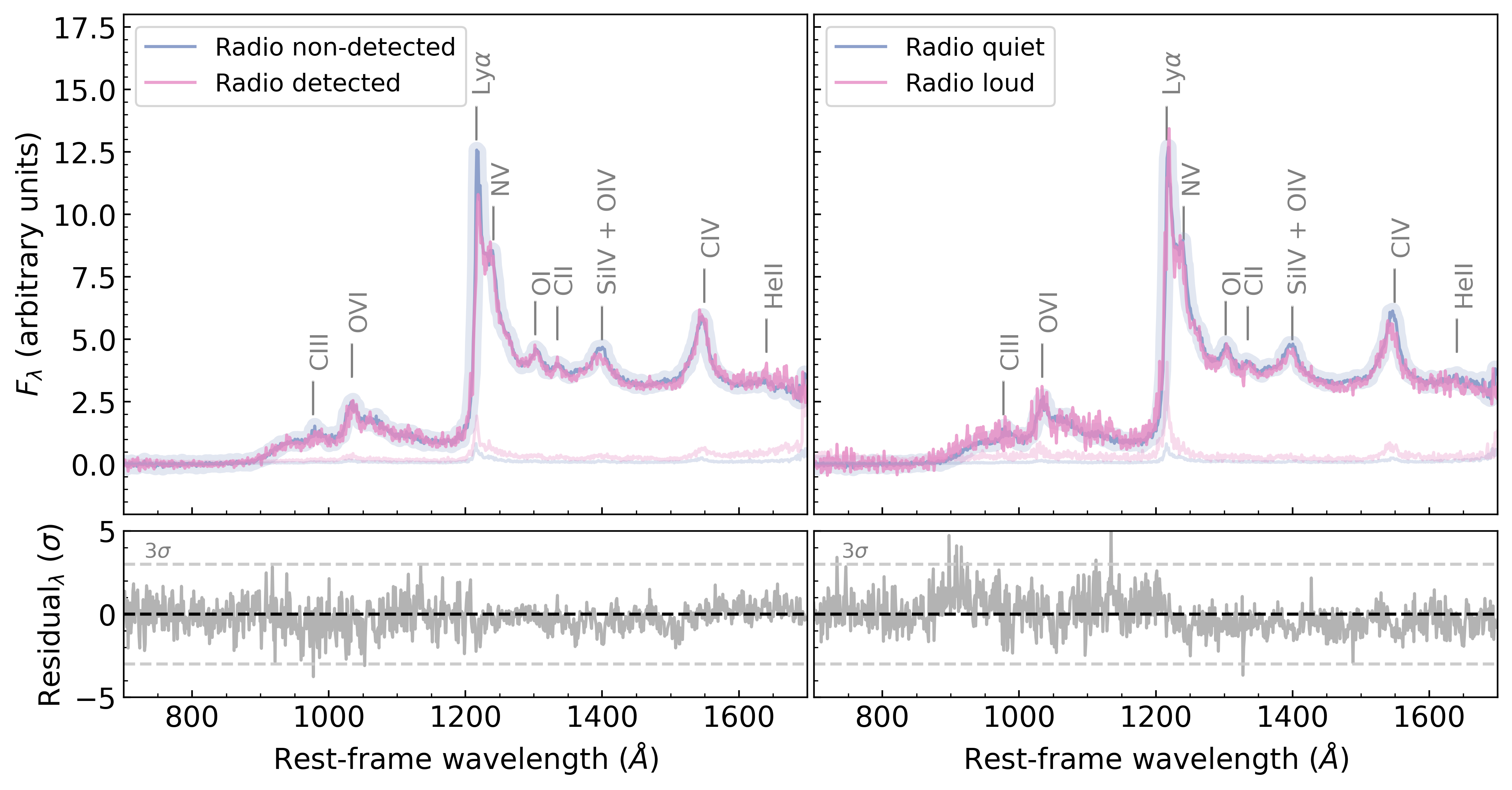}\vspace{-0.0cm}
    \caption{Median composite of radio detected versus non-detected quasars (left) and radio-loud versus radio-quiet (right) at $4.70\leq z\leq5.70$. The stacked spectra of the radio non-detected and radio-quiet quasars are highlighted with a thick shaded line for easy comparison. The distributions below display the corresponding error spectra. The bottom panels show the residual spectra after subtracting the radio non-detected (radio-quiet) composite from the radio detected (radio-loud) composite and dividing by their uncertainties added in quadrature. All deviations are within the 3$\sigma$ range and therefore not significant.} 
    \label{fig:spec_stacks}
\end{figure*}

To compare the average spectral shapes of the radio detected versus non detected quasars we performed stacking using the code \textsc{SpecStacker}\footnote{\url{https://github.com/m-arnaudova/SpecStacker}} developed by \cite{Arnaudova2024MNRAS.528.4547A} for SDSS quasar spectra. Due to the normalization process of the quasar spectra, the stacking needs to be performed on small redshift ranges to ensure a common wavelength grid. Due to the low number of quasars at $z>5.7$, we performed the stacking separately in two redshift bins, $4.7 < z \leq 5.0$ and $5.0 < z \leq 5.7$, which we then combine and rescale. The resulting stacks for both the radio detected versus non-detected, and radio-loud versus -quiet are shown in Fig.~\ref{fig:spec_stacks}. The lower panels show the residual in uncertainty units, which was calculated by subtracting the two stacked spectra and dividing the result by their combined uncertainties added in quadrature. These residuals demonstrate that the stacked spectra are generally within a 1$\sigma$ agreement, and never deviate more than 3$\sigma$. We note that the continuum redwards of Ly$\alpha$ is slightly weaker for the stack of radio-loud quasars, however, this difference is not significant. Therefore, there is no evidence for any difference in general rest-frame UV spectra between high-$z$ radio quasars and radio non-detected quasars. \cite{Arnaudova2024MNRAS.528.4547A} found evidence for radio-loud quasars at $0.6<z<3.4$ having a redder continuum and enhanced [O\textsc{ii}] emission than their radio quiet counterparts. However, a larger wavelength coverage at $>2000$ \AA\, rest-frame is needed to confirm this at higher redshift, which can only be achieved with near-infrared observations.

\section{Discussion}
\label{sec:discussion}

\subsection{Are radio bright high-z quasars special?}

The sample presented in this work represents the largest spectroscopic uniform sample of radio quasars at $z\sim5$ to date, which allowed for a direct comparison of the rest-frame UV spectral properties of radio-loud and radio-quiet, and radio detected and non-detected quasars. Our results suggest that generally their rest-frame UV spectral properties, such as emission line equivalent widths and C$\textsc{iv}$ blueshifts, are similar. The results obtained specifically for the WLQ and BAL phenomena will be discussed in Sect.~\ref{subsec:discussion_WLQs} and ~\ref{subsec:discussion_BALs}, respectively. 
One of the issues in comparing the rest-frame UV emission to the radio emission is that these generally probe different timescales. The rest-frame UV probes recent accretion, while low frequency radio traces jets that could already be hundreds of millions of years old. The current state of the accretion disk can therefore not be directly compared to its state during the launching of the jet in the case of these quasars. 

However, these radio jets are expected to influence both their immediate interstellar medium via AGN feedback and the surrounding intergalactic medium when the radio jets grow beyond their galaxy confinements (e.g., \citealt{Miley2008A&ARv..15...67M, Hardcastle2020NewAR..8801539H}). At high-$z$, the Universe is much denser with typical dark matter halos of $\sim10^{12}$ M$_{\odot}$ with a virial radius of only $r_{200c}\approx50$ kpc at $z\sim5$, the radius where the mean density is 200$\times$ higher than the mean cosmic density (e.g., \citealt{Harikane2022ApJS..259...20H}). Powerful radio jets, while rare, can therefore have an important impact on the evolution of their galaxy and their environment in the early Universe.

Not finding evidence of radio-loudness evolution (apart from potentially missing the tail at high-$z$, see Sect.~\ref{subsec:RL_evolution}) from $z\sim2$ to $z\sim5$ furthermore highlights that radio-loud quasars are already active within the first Gyr after the Big Bang. Probing quasars at higher redshifts with deeper radio observations, to probe the complete population, will be necessary to further study any radio-loudness evolution, which could indicate changes in intrinsic SMBH properties (e.g., accretion mode, black hole mass or spin) and environment (e.g., gas reservoir).  

Current observations suggest there is an evolution of extended radio jet sizes, with giant radio lobes ($\gtrsim$100 kpc) completely missing at $z>4$ while abundant at low-$z$ (e.g., \citealt{Fabian2014MNRAS.442L..81F, Gloudemans2025ApJ...980L...8G}). This could plausibly be an observational effect, since inverse Compton (IC) scattering of CMB photons increases significantly at high-$z$ with $\propto(1+z)^4$ \citep{Morabito2018MNRAS.480.2726M}, which causes the apparent dimming of radio emission from extended radio lobes. The current LoTSS-DR3 resolution of 6\arcsec\, is not sufficient to resolve the radio structure of these quasars and therefore future high-resolution long baseline radio imaging will be crucial to investigate any size evolution.

\subsection{Redshift evolution of WLQs and implications of their radio emission}
\label{subsec:discussion_WLQs}

In our sample, we found an overall WLQ fraction of $\sim5\%$ using the criteria EW$_{\text{Ly}\alpha} < 15.4$ \AA (see Sect.~\ref{subsec:results_WLQs}). This WLQ fraction is lower than previous high-$z$ quasar studies, which range from 10-38\%, and suggested potential redshift evolution of WLQs (e.g.,~\citealt{Banados2014AJ....148...14B, Banados2016ApJS..227...11B, Shen2019ApJ...873...35S, Gloudemans2022A&A...668A..27G}). Selection effects could contribute to these varying fractions of WLQs. In principle, WLQs are more difficult to select at low-$z$ than high-$z$, because at low-$z$ there is no Lyman break to identify the quasar. More generally, the color selection of WLQs depends on both emission-line strength and redshift, with weak emission lines affecting the photometry used for quasar selection (e.g., \citealt{Diamond-Stanic2009ApJ...699..782D}).
Furthermore, luminosity-dependent effects such as the Baldwin effect (the anti-correlation between \ion{C}{4} EW and luminosity; \citealt{Baldwin1977ApJ...214..679B}) may contribute to this trend, since high-$z$ quasar samples are typically more luminous on average. 
Any potential evolution observed in WLQs could therefore be due to selection effects. In J. Yang et al. (in prep.), the WLQ fractions will be determined based on the final quasar sample, including a correction for selection bias. 

Furthermore, while this work finds an increased fraction of WLQs among radio-loud and radio-detected quasars, the difference is not significant ($\sim1.3\sigma$), so a larger sample is needed to draw firmer conclusions. We also did not find a significant correlation between radio-loudness and either \ion{C}{4} or Ly$\alpha$ EW.
Earlier work by \cite{Plotkin2015ApJ...805..123P} suggested it was unlikely that jet emission caused the weak broad emission lines in their sample, while \cite{Meusinger2014A&A...568A.114M} found WLQs to be 5$\times$ more likely to be radio detected than normal quasars, although with a lower mean radio-loudness. In this work, the percentage of Ly$\alpha$ (\ion{C}{4}) WLQs increased from 3.8$\pm$1.1\% (3.5$\pm$1.3\%) for radio non-detected to 9.5$\pm$4.2 (11.4$\pm$5.7\%) for radio-detected quasars. 
Radio-quiet quasars, not only radio-loud quasars, are known to host (low-luminosity) radio jets (e.g., \citealt{Leipski2006A&A...455..161L}), which could also contribute to boosting of the UV continuum and diluting of the line strength. The WLQ phenomenon has also been hypothesized to be caused by an undeveloped BLR or high Eddington accretion rate (e.g., \citealt{Leighly2007ApJS..173....1L, Meusinger2014A&A...568A.114M}). Near-infrared follow-up observations are necessary to investigate this scenario for the quasar sample in this work, and we again note that an increase in sample size is vital to verify the highly uncertain trends observed in this work.

Finally, we caution that the EW measurements depend heavily on the continuum fit. At high-$z$, the continuum is challenging to constrain, since optical surveys like DESI cover only wavelengths up to 1637 \AA\, at $z=5$. Near-infrared spectroscopy is therefore necessary to improve the constraints on the continuum and reliability of the EW measurements at high redshift.

\subsection{Redshift evolution of BAL quasars} 
\label{subsec:discussion_BALs}

As presented in Sect.~\ref{subsec:results_BALs}, our findings suggest that the BAL quasars in our high-$z$ sample are similarly likely to be radio detected than normal quasars within the expected errors. The fraction of radio detected BAL quasars potentially increases with \ion{C}{4} absorption index. However, most BAL quasars are radio-quiet, and there is no apparent correlation between radio-loudness and \ion{C}{4} absorption index (see Fig.~\ref{fig:RL_BAL_QSO_combined}). These results are generally consistent with the findings of \cite{Morabito2019A&A...622A..15M} for a large sample of quasars at $1.7 \leq z \leq 4.3$ using the balnicity index. \cite{Knigge2008MNRAS.386.1426K} demonstrated that while the absorption index yields a higher completeness of BAL quasars, it also decreases the accuracy of BAL identification. Using the BI instead of AI, only 63 quasars in our $z\leq5.28$ sample are identified as BAL quasars ($\sim13\%$) with only 11 of those detected in LoTSS-DR3. Also in this case, there is no correlation between BI and radio-loudness (Spearman's correlation coefficient = 0.045, $p$-value=0.89). The BAL quasars not selected by BI do demonstrate clear absorption features in their spectra by visual inspection. 

As suggested by \cite{Morabito2019A&A...622A..15M}, these findings indicate that radio emission and the BAL phenomenon are potentially linked to the same underlying process, which is already in place by $z\sim5$ as demonstrated here. In the evolutionary picture, where young quasars produce jets and begin to drive outflows that are eventually cleared, we might expect the BAL fraction to increase at high-$z$. Recent work by \cite{Bischetti2022Natur.605..244B, Bischetti2023ApJ...952...44B} indeed found that the BAL fraction appears constant at $z\lesssim4.5$ while increasing by a factor of $\sim2-3$ at $z\gtrsim6$ to almost 50\%. In our sample at $4.7\lesssim z \lesssim 6.7$, the BI fraction found of 13\% is similar to the 8-15\% observed BAL fraction found for UV-selected quasars at $2\lesssim z\lesssim4$ by \cite{Reichard2003AJ....126.2594R}, \cite{Knigge2008MNRAS.386.1426K}, \cite{Gibson2009ApJ...692..758G}, \cite{Allen2011MNRAS.410..860A}, and \cite{Morabito2019A&A...622A..15M}. The final DESI $z\gtrsim5$ quasar sample in future work will be used to measure the intrinsic fraction of BAL quasars by correcting for selection effects. 
As discussed in Sect.~\ref{sec:intro}, the orientation of quasars can also play an important role in the BAL phenomenon. The radio-bright BALs identified in this work are ideal candidates for high-resolution radio follow-up to constrain radio jet sizes and radio spectral analysis, which can provide necessary evidence for the orientation versus evolution model, since these can be used as proxies for jet age and orientation.

\section{Summary}
\label{sec:summary}

By combining new quasars at $4.7\leq z \leq 6.8$ discovered by a dedicated DESI target program with the LOFAR Two Metre Sky Survey, we built the largest uniform spectroscopic sample of radio-detected quasars into the cosmic dawn to date. Using a forced Gaussian fitting technique, we identified 83 radio-emitting quasars out of the 510 quasars covered by LoTSS. With high radio luminosities of $L_{144\text{MHz}}= 3\times10^{25} - 3\times10^{28}$ W Hz$^{-1}$, their radio emission is likely dominated by jets, however, winds and shocks can also contribute to the produced radio emission. 

Using the classical definition of radio-loudness, we find that 5-15\% of the quasars in our parent sample can be considered radio-loud and 85-95\% radio-quiet, with the range caused by ambiguous quasars with $3\sigma$ upper limits above the traditional $R=10$ threshold. Most of these ambiguous radio-loud/radio-quiet quasars are located at Dec.$<10^{\circ}$ where the sensitivity of LoTSS-DR3 decreases. Considering only the quasars with Dec.$>10^{\circ}$ results in a more constrained radio-loud fraction of $\sim6-9$\%. We performed statistical tests across three sample pairs: radio-loud versus radio-quiet, their ``pure'' subsamples restricted to Dec.$>10^{\circ}$, and radio-detected versus non-detected quasars. Radio-loud/radio-quiet classification for ambiguous sources is determined via Monte Carlo realizations. Our (spectral) analysis of these quasar samples resulted in the following: 

\begin{itemize}
\itemsep0em 
    \item Comparing our $z\sim5$ radio detected quasar sample to a luminosity-matched sample at $z\sim2$, we did not find any evidence for radio-loudness evolution. This suggests that radio jets are already well-established in quasars at $z\sim5$. The high-$z$ sample does lack sources in the high radio-luminosity tail, however, this could be due to low number statistics. 
    
    \item The Ly$\alpha$ and \ion{C}{4} EW distributions differ significantly between radio-detected and non-detected quasars —radio-detected sources show on average smaller EWs— while no significant differences are found between radio-loud and radio-quiet samples (including the pure subsamples), possibly due to limited statistics. 
    The fractions of WLQs are higher in all radio-loud and radio-detected samples compared to radio-quiet and non-detected. However, due to the limited number statistics, this difference corresponds to only $\sim1.3\sigma$ for radio detected versus non-detected and $\sim1\sigma$ for (pure) radio-loud versus radio-quiet, and therefore a larger sample is required to confirm an increased WLQ fraction among high-$z$ radio quasars.
    
    \item We identified 106 BAL quasars in our sample, which are predominantly radio-quiet. We found no correlation between radio-loudness and absorption index, implying that the BAL outflows themselves do not produce the radio emission. This is consistent with previous works at low-$z$. More investigation is needed to determine whether young quasar age or orientation is the predominant cause of the broad absorption lines.
    
    \item The distribution of radio-loud quasars in \ion{C}{4} velocity shift is similar to the distribution of radio-quiet quasars, without any significant correlation with radio-loudness. We did not find a skew towards lower \ion{C}{4} blueshift for radio-loud quasars as previously found at low-$z$, however, this could be due to our small sample size or to the different blueshift methodology due to the lack of systemic redshifts for our quasars.
    
    \item Comparison of the composite rest-frame UV spectra of all quasar samples yielded no evidence of any differences in spectral features and continuum of radio-detected and radio-loud quasars at high-$z$. To provide a direct comparison to low-$z$ quasar samples, (near-)infrared wavelength coverage is necessary.  
\end{itemize}

This work highlights that radio jets are already well-established at $z\sim5$ with little or no redshift evolution in their rest-frame UV spectral properties. However, even with this relatively high number of radio quasars, we are still dealing with low-number statistics, which hinders drawing any firm conclusions. Only 6 quasars in the sample show possible extended radio features, highlighting the necessity of high-resolution radio follow-up to disentangle core and jet emission. 

In the near future, the \textit{SKA} telescope is expected to be able to detect fainter populations of radio quasars and galaxies out to high-$z$, which will revolutionize this field. Furthermore, the discovery of quasars out to $z\sim9$ from the \textit{Euclid} wide survey \citep{Euclid2019A&A...631A..85E} will enable tracing any evolution to the earliest epochs.

\begin{acknowledgments}

We thank the anonymous referee for a careful reading of the manuscript and for constructive comments that improved the paper. We are grateful to Ting-Yun Cheng, Emanuele Paolo Farina, Malgorazata Siudek, and Benjamin Weaver for their review and helpful feedback. Furthermore, we thank Jiaxi Yu for their guidance during the publication process. 

DJBS acknowledges support from United Kingdom’s Science and Technology Facilities Council (STFC) under grant ST/V000624/1, and from the Leverhulme Trust via Research Project Grant RPG-2025-078.
LKM is grateful for support from a UKRI FLF [MR/Y020405/1] and LOFAR-UK via STFC [ST/V002406/1].
SP is supported by the international Gemini Observatory, a program of NSF NOIRLab, which is managed by the Association of Universities for Research in Astronomy (AURA) under a cooperative agreement with the U.S. National Science Foundation, on behalf of the Gemini partnership of Argentina, Brazil, Canada, Chile, the Republic of Korea, and the United States of America.
VAF acknowledges support from United Kingdom Research and Innovation (grant code: MR/V022830/1).

This material is based upon work supported by the U.S. Department of Energy (DOE), Office of Science, Office of High-Energy Physics, under Contract No. DE–AC02–05CH11231, and by the National Energy Research Scientific Computing Center, a DOE Office of Science User Facility under the same contract. Additional support for DESI was provided by the U.S. National Science Foundation (NSF), Division of Astronomical Sciences under Contract No. AST-0950945 to the NSF’s National Optical-Infrared Astronomy Research Laboratory; the Science and Technology Facilities Council of the United Kingdom; the Gordon and Betty Moore Foundation; the Heising-Simons Foundation; the French Alternative Energies and Atomic Energy Commission (CEA); the National Council of Humanities, Science and Technology of Mexico (CONAHCYT); the Ministry of Science, Innovation and Universities of Spain (MICIU/AEI/10.13039/501100011033), and by the DESI Member Institutions: \url{https://www.desi.lbl.gov/collaborating-institutions}. Any opinions, findings, and conclusions or recommendations expressed in this material are those of the author(s) and do not necessarily reflect the views of the U. S. National Science Foundation, the U. S. Department of Energy, or any of the listed funding agencies.

The authors are honored to be permitted to conduct scientific research on I'oligam Du'ag (Kitt Peak), a mountain with particular significance to the Tohono O’odham Nation.

LOFAR is the Low Frequency Array designed and constructed
by ASTRON. It has observing, data processing, and data storage facilities in several countries, which are owned by various
parties (each with their own funding sources), and which are
collectively operated by the ILT foundation under a joint scientific policy. The ILT resources have benefited from the following recent major funding sources: CNRS-INSU, Observatoire de
Paris and Université d’Orléans, France; BMBF, MIWF-NRW,
MPG, Germany; Science Foundation Ireland (SFI), Department
of Business, Enterprise and Innovation (DBEI), Ireland; NWO,
The Netherlands; The Science and Technology Facilities Council, UK; Ministry of Science and Higher Education, Poland; The
Istituto Nazionale di Astrofisica (INAF), Italy

\end{acknowledgments}

\begin{contribution}

The first-tier DESI authors contributed significantly to the target program, the data analysis, and science presented in this manuscript. 
Second-tier DESI authors, the DESI builders, are recognized for their significant contributions to the data products used in this work.

All authors from the LOFAR collaboration played either leadership or significant supporting roles in one or more of: the management and development of the International LOFAR Telescope, its software data pipelines, the creation of science-ready data products, and the preparation of this manuscript. 

\end{contribution}

\facilities{Mayall (DESI), Mayall (Mosaic-3), Blanco (DECam), Astro Data Lab, LOFAR, VLA, WISE, NEOWISE}

\software{Astropy \citep{astropy+2013, astropy+2018, astropy+2022}, EAZY \citep{brammer2011ApJ...739...24B}, Matplotlib \citep{matplotlib}, NumPy \citep{numpy}, PyQSOFit \citep{Guo2018ascl.soft09008G}, QuasarNet \citep{Busca2018arXiv180809955B}, Redrock \citep{Bailey2026ascl.soft06024B}, SPARCL \citep{sparcl}}

\section{Data availability}
The data used to make the plots in this paper can be found at \url{https://doi.org/10.5281/zenodo.21727117}.

\appendix

\section{Radio images \& optical overlays}
\label{sec:appendix_radio}

\begin{figure*}
    \centering
    \includegraphics[width=0.85\textwidth, trim={0.0cm 0.0cm 0.0cm 0.0cm}, clip]{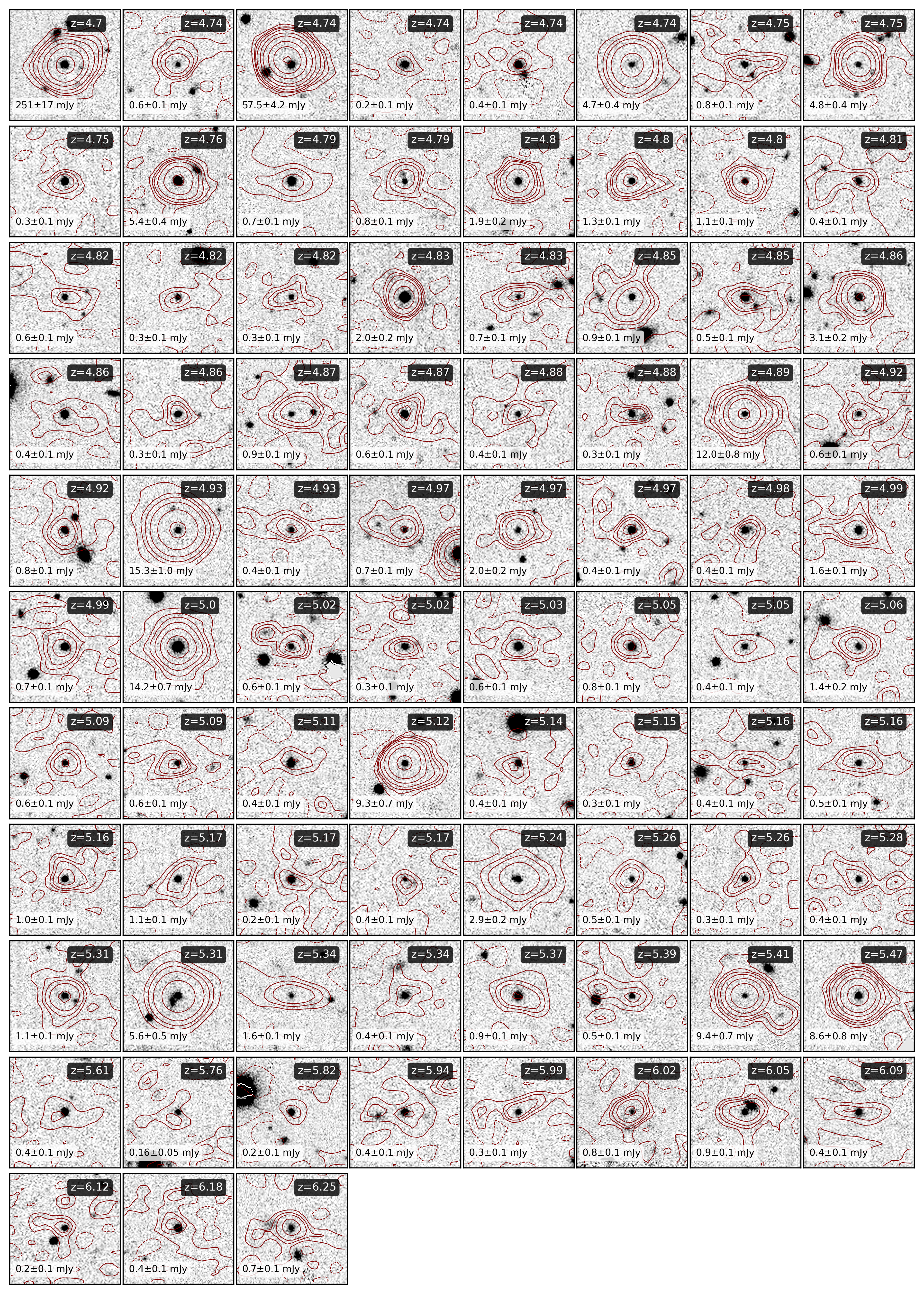}\vspace{-0.0cm}
    \caption{Radio detections 83 DESI quasars at $z\geq4.7$ sorted on redshift. The radio contours are drawn at [$1,2,3,5,10,20,50,100,200$]$\times$rms and superimposed on optical z-band images of the DESI Legacy Imaging Survey DR8 (30\arcsec$\times$30\arcsec\, cutout). The dashed red lines show the negative $-1\times$rms contours. The total flux density of each source at 144 MHz is given in the bottom left corner of each panel.} 
    \label{fig:detected_radio_sources}
\end{figure*}

\begin{figure*}
    \centering
    \includegraphics[width=1.0\textwidth, trim={0.0cm 0.0cm 0.0cm 0.0cm}, clip]{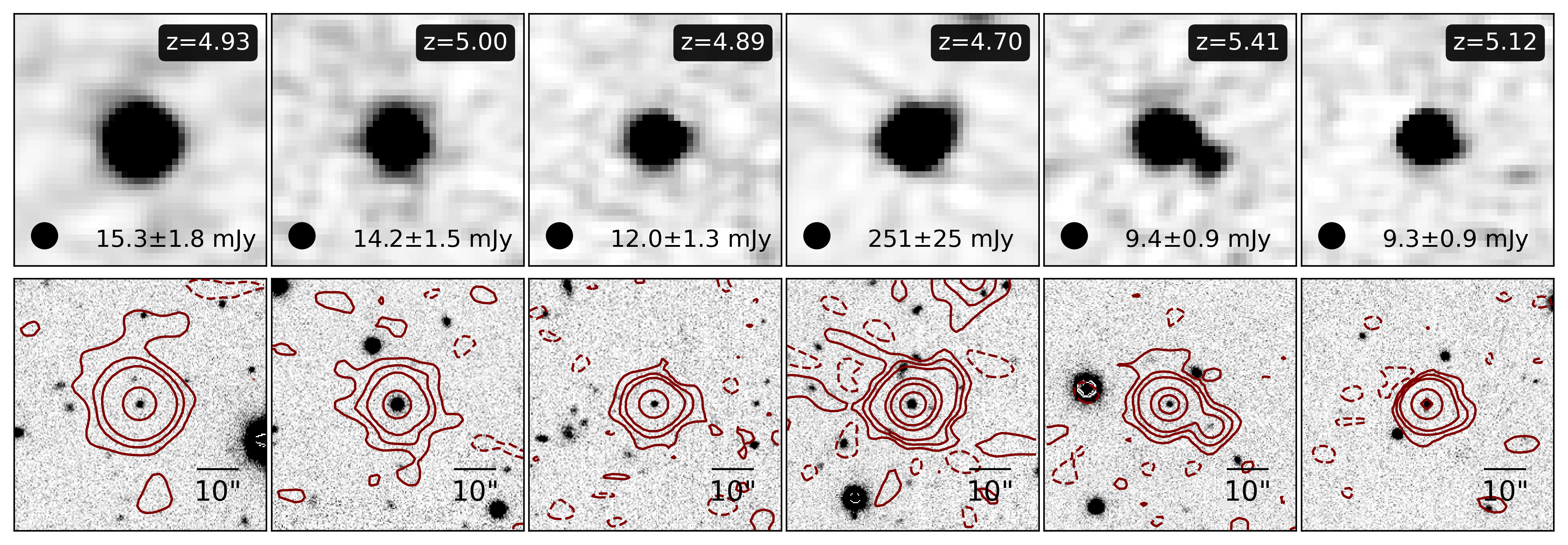}
    \caption{Radio quasars identified as extended in the LoTSS-DR3 survey. The top panels show their LoTSS 144 MHz images. The bottom panels show their radio emission superimposed on optical z-band images (60\arcsec$\times$60\arcsec\, cutout). The radio contours (solid red lines) are drawn at [$2, 5, 10, 50, 100, 250, 500$]$\times$rms. The red dashed lines show the negative $-2\times$rms contours. The second radio component apparent in the fifth panel is likely associated with the galaxy detected in optical.}
    \label{fig:extended_radio_sources}
\end{figure*}

Figure \ref{fig:detected_radio_sources} shows the LoTSS-DR3 detections of all 83 DESI quasars. The radio contours are superimposed on Legacy z-band images. While some radio emission is close to the sensitivity limit, the association with the optical quasar is convincing. Figure \ref{fig:extended_radio_sources} shows the six quasars with extended radio emission. The extended radio emission of DESI J134748+440546 (5th panel) is likely caused by the nearby galaxy visible in the z-band image. 

\section{PyQSOFit spectra fitting examples}
\label{sec:appendix_pyqsofit}

\begin{figure*}
    \centering
    \includegraphics[width=0.95\textwidth, trim={0.0cm 0.0cm 0.0cm 0.0cm}, clip]{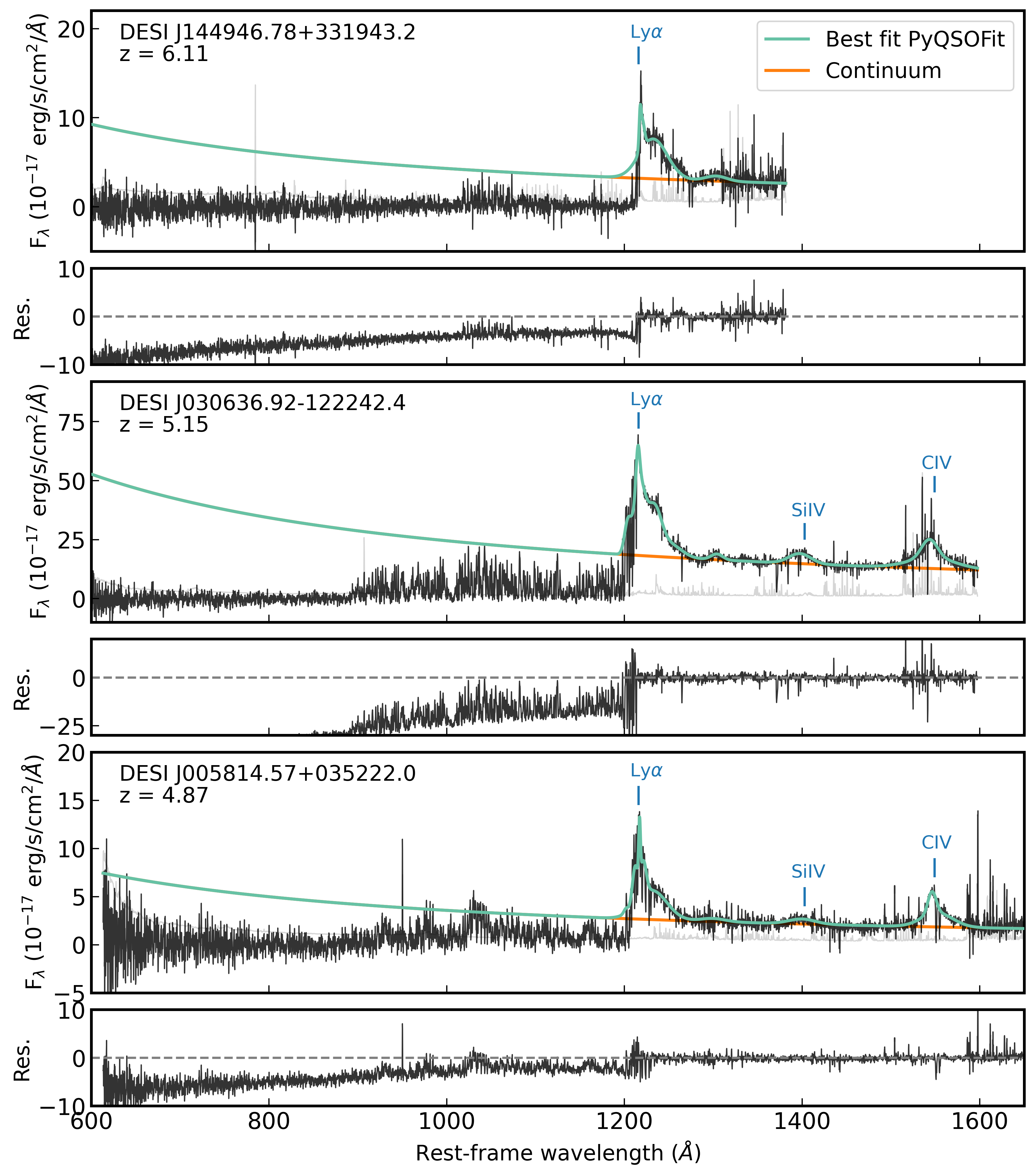}
    \caption{Example spectra (black) and best fits (green) of three quasars at different redshifts using the PyQSOFit fitting routine as described in Sect.~\ref{subsec:pyqsofit}. These quasar spectra were previously published in \cite{Yang2023ApJS..269...27Y}. The continuum fit is shown in orange for a slope of $\beta = -1.5$. The panels below each spectrum show the residual spectrum after subtracting the model.} 
    \label{fig:pyqsofit_examples}
\end{figure*}

Figure \ref{fig:pyqsofit_examples} shows example spectra of three high-$z$ quasars (DESI J144946.78+331943.2, DESI J030636.92-122242.4, DESI J005814.57+035222.0) at different redshifts, which were previously published in \cite{Yang2023ApJS..269...27Y}. The figure demonstrates the performance of our PyQSOFit routine as described in Sect.~\ref{subsec:pyqsofit}, which resulted in reduced $\chi^2$ values of the Ly$\alpha$ and \ion{C}{4} emission line fits close to 1.  

\bibliography{main}{}
\bibliographystyle{aasjournalv7}




\end{document}